\documentclass{IJMRmodified}
\usepackage{a4}
\usepackage{caption}

\newif\ifdraft

\usepackage{amssymb}
\usepackage{array}
\usepackage[english]{babel}
\usepackage{color}
\usepackage{colortbl}
\usepackage{dcolumn}
\usepackage{graphicx}
\usepackage{hyperref}
\usepackage[utf8]{inputenc}
\usepackage{latexsym}
\usepackage{multicol}
\usepackage{multirow}
\usepackage{soul}
\usepackage{pifont}
\usepackage{rotating}
\usepackage{url}
\usepackage{wrapfig}

\newcommand{\blue}[1]{\ifdraft{\leavevmode\color{blue}{#1}}\else{\leavevmode\color{black}{#1}}\fi}

\newcommand{\side}[1]{\multicolumn{1}{c|}{\begin{sideways}#1\end{sideways}}}

\newtheorem{revcomment}{Reviewer's Comment \#}

\begin{document}


\title{Building Automated Survey Coders \\ via Interactive Machine
Learning}

\author{Andrea Esuli, Alejandro Moreo, Fabrizio Sebastiani \thanks{The
order in which the authors are listed is purely alphabetical; each
author has given an equally
important contribution to this work.} \\
Istituto di Scienza e Tecnologie dell'Informazione \\
Consiglio Nazionale delle Ricerche \\ 56124 Pisa, Italy \\ Email:
\textsf{firstname.lastname}@isti.cnr.it}

\begin{abstract}
  \noindent Software systems trained via machine learning to
  automatically classify open-ended answers (a.k.a.\ \emph{verbatims})
  are by now a reality. Still, their adoption in the survey coding
  industry has been less widespread than it might have been. Among the
  factors that have hindered a more massive takeup of this technology
  are the effort involved in manually coding a sufficient amount of
  training data, the fact that small studies do not seem to justify
  this effort, and the fact that the process needs to be repeated anew
  when brand new coding tasks arise. In this paper we will argue for
  an approach to building verbatim classifiers that we will call
  ``Interactive Learning'', and that addresses all the above
  problems. We will show that, for the same amount of training effort,
  interactive learning
  delivers much better coding accuracy than standard
  ``non-interactive'' learning.
  This is especially true when the amount of data we are willing to
  manually code is small, which makes this approach attractive also
  for small-scale studies.  Interactive learning also lends itself to
  reusing previously trained classifiers for dealing with new (albeit
  related) coding tasks.  Interactive learning also integrates better
  in the daily workflow of the survey specialist, and delivers a
  better user experience overall.
\end{abstract}

\maketitle

\noindent {\footnotesize \textbf{Keywords}}. Machine Learning;
Automated Survey Coding; Artificial Intelligence; Verbatim Coding;
Sentiment Classification.


\section{Introduction}
\label{sec:intro}

\noindent In fields such as market research, the social sciences,
political science, and customer relationship management, data are
often collected through surveys, conducted by a survey specialist and
involving a number of respondents \cite{de-Vaus:2014jw}. Conducting a
survey usually involves a questionnaire, i.e., a list of questions
which respondents are asked to answer. The majority of questions to be
found in questionnaires are of the ``closed'' type, where the
respondent is required to tick one of a predefined set of
answers. \emph{Open} (a.k.a.\ ``open-ended'') \emph{questions} instead
involve returning a textual answer, whose length is not specified a
priori. When computing the results of the survey, closed answers and
open answers require very different amounts of processing: while
closed answers simply involve checking which (or how many) respondents
have picked which predefined options, open answers require more
complex analysis. In order to manage open answers, the survey
specialist first defines a classification scheme, i.e., a set of
classes of interest for the given application (e.g.,
`BadCustomerSupport', `IssuesWithWebsite', etc., for a customer
satisfaction survey run by a telecom company), and then classifies
(i.e., attributes one or more classes from the classification scheme
to) each answer based on its textual content. The results of the
survey are then obtained by checking which (or how many) respondents'
answers have been attributed which class.

In the language of survey specialists, classes are called
\emph{codes}; classification schemes are called \emph{codeframes}, or
\emph{codebooks}; classification is called \emph{coding}; human
annotators are called \emph{coders}; the answers returned to
open-ended questions are called \emph{verbatims}; and the task of
classifying open-ended answers is called \emph{survey coding}, or
\emph{verbatim coding}. This is the terminology we will adopt in the
rest of the paper.

Open questions have advantages and disadvantages with respect to their
closed counterparts. On the plus side, it is generally acknowledged
that answers returned to open questions are richer and more
informative, since the respondent can express her thoughts more
freely, not being constrained by \textit{a priori} choices made by the
survey specialist. On the other hand, managing open questions is more
onerous, since manually coding them is a time-consuming task.


\subsection{The Quest for Automated Survey Coding}
\label{sec:surveycoding}

\noindent In the attempt to make open questions more manageable for
survey specialists, a number of researchers have proposed using
software systems for coding open answers automatically.  This paper
looks at a specific type of survey coding systems, namely, those based
on \emph{machine learning} (ML -- see \cite{Murphy:2012eu}).  ML, a
subfield of \emph{artificial intelligence} (AI), has recently taken
the IT world by storm, to the point that Andrej Karpathy, Director of
AI at Tesla, has described neural networks (the current protagonists
of ML) as representing ``the beginning of a fundamental shift in how
we write software''.\footnote{Andrej Karpathy, `Software 2.0',
\emph{Medium}, November 11, 2017,
\url{https://medium.com/@karpathy/software-2-0-a64152b37c35}}

The most important form of ML is \emph{supervised learning}, according
to which a learning algorithm ``trains'' a software system to perform
a certain task by showing it a number of correctly solved instances
(called \emph{training examples}) of this task; by analysing these
instances and their correct solutions the system learns to solve new
instances itself. This is called ``supervised'' learning, because the
human operator who feeds the training examples to the learning
algorithm plays the role of the ``supervisor''. \blue{Among the tasks
that can be solved via supervised ML, classification is certainly the
most important, since many real-world problems that involve prediction
of future or unknown events can be framed as classification problems.}

The world of survey coding has not been immune to the ML revolution.
ML-based software systems trained to automatically code verbatims are
by now a reality; they have been described in a number of publications
(see e.g.,
\cite{Clarke:2011os,Esuli:2010kx,Gamon:2004:SCC,JASIST03,ASCIC03,MRS07,Patil:2013zp}),
and are offered by several commercial vendors.  A ML-based survey
coding system learns, from sample manually coded verbatims (the
training examples), the characteristics a new uncoded verbatim should
have in order to be attributed a given code. The ML approach to
building survey coding systems is advantageous with respect to the
more traditional ``rule-based'' approach \cite{Viechnicki98}, since it
requires much less humanpower (see \cite{Esuli:2010kx} for a thorough
discussion of this point).

However, notwithstanding their attractive properties, the adoption of
ML-based systems in the survey coding industry has been less
widespread than it might have been. Several factors have hindered a
more massive takeup of this technology. One of these factors is the
effort involved in manually coding an amount of training data
sufficient to guarantee a good enough coding accuracy on the part of
the trained system. A second reason is the fact that small studies
(i.e., coding tasks in which the amount of verbatims that require
coding is small) do not seem to justify this effort, since the amount
of uncoded verbatims that need to be manually coded for use as
training data is, for small studies, close to the size of the study
itself.  Yet another reason is the fact that the process needs to be
repeated anew when brand new coding tasks arise; in other words, the
training data generated for coding a given study cannot be reused when
a new, different study comes up, unless the two studies share the same
codeframe \emph{and} consist of data from the same source.


\subsection{What this Paper is about}
\label{sec:about}

\noindent In this paper we argue for an approach to building verbatim
classifiers which addresses the above problems, and which we call
``Interactive Learning''.  We show that, for the same amount of
training effort, interactive learning delivers substantially better
coding accuracy than standard ``non-interactive'' learning. This is
especially true when the amount of data we are willing to manually
code for training purposes is small, which makes this approach
attractive also for small-scale studies.  Interactive learning also
lends itself to reusing previously trained classifiers for dealing
with new (albeit related) coding tasks. Interactive learning also
integrates better in the daily workflow of the user, and delivers a
better user experience overall.

The rest of the paper is organized as follows. Section
\ref{sec:interactive} discusses the interactive learning approach to
survey coding and its rationale. Section \ref{sec:experiments}
presents the results of a number of experiments on various datasets of
open-ended answers, in which we compare the accuracy and the
efficiency of a verbatim coding system built via interactive learning,
with those of a more traditional machine-learned system.  Section
\ref{sec:reuse} looks, with the support of experimental data, at how
systems built via interactive learning lend themselves to ``classifier
reuse'', i.e., to leveraging, for solving a given coding task, a
classifier previously trained for a different but related coding task.
In Section \ref{sec:related} we look at related work in the area of
automated verbatim coding, and discuss differences between our
approach and other published work.  Section \ref{sec:discussion} sums
up our discussion, pointing at avenues for further development.


\section{Interactive Machine Learning for Automated Verbatim Coding}
\label{sec:interactive}

\noindent \blue{Before discussing interactive learning we define some
standard machine learning terminology for classification tasks.

There are different classification problems of applicative interest,
based (a) on how many classes the codeframe $\mathcal{C}$ contains,
and (b) on how many of the classes in $\mathcal{C}$ can be
legitimately attributed to the same item (in our case: to the same
verbatim). Both (a) and (b) are not designer choices, but are imposed
by the application. 
We characterize classification problems as follows:
\begin{enumerate}
\item \emph{Single-Label} classification is defined as classification
  when each item must belong to exactly one of the classes in
  $\mathcal{C}=\{c_{1}, ..., c_{m}\}$.
\item \emph{Multi-Label} classification is defined as classification
  when the same item may belong to any number of classes (zero, one,
  or several) in $\mathcal{C}=\{c_{1}, ..., c_{m}\}$.
\item \emph{Binary} classification may alternatively be defined
  \begin{enumerate}
  \item \label{item:BinAsSLc} as single-label classification with
    $m=2$ (in this case $\mathcal{C}=\{c_{1},c_{2}\}$ and
    each item must belong to either $c_{1}$ or $c_{2}$).
  \item \label{item:BinAsMLc} as multi-label classification with
    $m=1$ (in this case $\mathcal{C}=\{c\}$ and each item
    either belongs or does not belong to $c$).
  \end{enumerate}
\item To distinguish it from binary classification, single-label
  classification with $m>2$ is called \emph{Single-Label
  Multi-Class} (SLMC) classification.
\item To distinguish it from binary classification, multi-label
  classification with $m>1$ is called \emph{Multi-Label
  Multi-Class} (MLMC) classification.
\end{enumerate}


\noindent MLMC classification can be reduced to (i.e., solved in terms
of) binary classification. In fact, one may trivially solve a MLMC
problem by independently training $m$ binary classifiers, one for each
code $c_{i}$ in $\mathcal{C}$. Once trained, the binary classifier for
code $c_{i}$ will be entrusted with the task of deciding whether
$c_{i}$ applies to item $d$ or not. So, by running the $m$ binary
classifiers (conceptually) in parallel, zero, one, or several codes at
the same time can be assigned to $d$. The classifier for code $c_{i}$
is trained by using the training examples labelled by $c_{i}$ as the
``positive training examples'', and the training examples \emph{not}
labelled by $c_{i}$ as the ``negative training examples''.  The SLMC
case cannot instead be recast into the binary case or into the MLMC
case.

For reasons that will be made clear in Section \ref{sec:experiments},
from here on we will assume we are dealing with binary
classification. Given what we have said above, this also implicitly
addresses MLMC classification.}

In order to explain the notion of ``interactive learning'', we now
move to discussing two dichotomies well-known in the field of machine
learning.


\subsection{Active Learning or Passive Learning?}
\label{sec:activevspassive}

\noindent The first dichotomy we illustrate is the one between
\emph{active learning} (AL) and \emph{passive learning} (PL), and
refers to the role that the system plays in the choice of the items
that should be manually coded, in order for them to play the role of
training data. ``Standard'' machine-learning-based verbatim coding
systems (such as, e.g., the one of \cite{ASCIC03}) rely on PL, since
it is the user who chooses the examples that, once manually coded,
will be used as training data. Instead, in AL it is the system that
provides the user with the verbatims to manually code. Existing AL
algorithms differ in terms of the policies they adopt for choosing the
items that the user should manually code; finding the policies that,
for a certain \emph{annotation budget} (i.e., the number of items that
the user is willing -- or paid -- to manually code), maximise the
accuracy of the resulting classifier, is the main goal of AL as a
discipline \cite{Cohn:2011lk}. Some such policies allow providing the
user with ``artificial'' (i.e., completely made up) uncoded items
(this is called \emph{constructive active learning}). However, the
by-now most frequently used approach to AL is \emph{pool-based} AL,
according to which the system chooses the items that the user should
manually code from a ``pool'' of available (and non-artificial)
uncoded items. Rather than choosing a subset of the items in the pool,
most pool-based AL algorithms \emph{rank} the items in the pool,
implying that the user should start coding from the top of the ranked
list and proceed down the ranking until the annotation budget is
over. Ranking all the elements in the pool, rather than choosing a
subset thereof, has the advantage that the available annotation budget
does not need to be determined in advance.

Rather than for training a classifier from scratch, AL is often used
for \emph{improving} an existing classifier originally trained via PL
(see e.g., \cite{Esuli:2010kx}), by asking the user to ``validate''
(i.e., manually code, thereby confirming or disconfirming the class
provisionally assigned to) a number of automatically coded
(``autocoded'') items. In this paper we will explore a different
avenue, i.e., \emph{one in which AL is used right from the beginning,
and in which PL thus plays no role.}


\subsection{Batch Learning or Incremental Learning?}
\label{sec:batchvsincremental}

\noindent The second dichotomy we touch upon is that of \emph{batch
learning} (BL) vs.\ \emph{incremental learning} (IL --
a.k.a. \emph{online learning}), and refers to the way the training
items are provided to the learning algorithm. In BL the training items
are provided to the learning algorithm all at the same time, and the
algorithm generates a trained classifier after analysing them
collectively. In IL \cite{Auer:2011lk} the training items are instead
provided to the learning algorithm one at a time. One application
scenario where this is useful is, e.g., when the training items are
not all available right from the beginning, and instead become
available over time.

In IL the input to the learning algorithm is \emph{not} a set of
training items, but a previously trained classifier plus one single
training item. What the algorithm does is ``update'' the existing
classifier by bringing to bear the information obtained from the new
training item; training a classifier is thus accomplished in a
step-by-step fashion, by carrying out as many classifier update
operations as there are training items.

Unlike in the AL vs.\ PL dichotomy, where the same learning algorithm
could be trained either via AL or via PL, the BL vs.\ IL dichotomy
translates into a sharp distinction between (a) algorithms that only
handle BL (which thus take as input a set of training items) and (b)
algorithms that only handle IL (which thus take as input a previously
trained classifier and a single training item).


\subsection{Interactive learning!}
\label{sec:interactivelearning}

\noindent In this paper we argue for an approach to building verbatim
classifiers based on ``interactive learning''. So, where does
interactive learning stand with respect to the two dichotomies
illustrated in Sections \ref{sec:activevspassive} and
\ref{sec:batchvsincremental}? Simply stated,

\begin{center}\textbf{Interactive Learning \\ = \\ Active Learning +
  Incremental Learning}\end{center}

\noindent With respect to the standard way machine-learning-based
verbatim coders are built, interactive learning thus takes opposite
stands with respect to both dichotomies; while the former are built
via batch passive learning, we now argue for systems built via a
combination of active learning and incremental learning.

\begin{figure}[tbh]
  \begin{center}
    \includegraphics[width=1.4\textwidth]{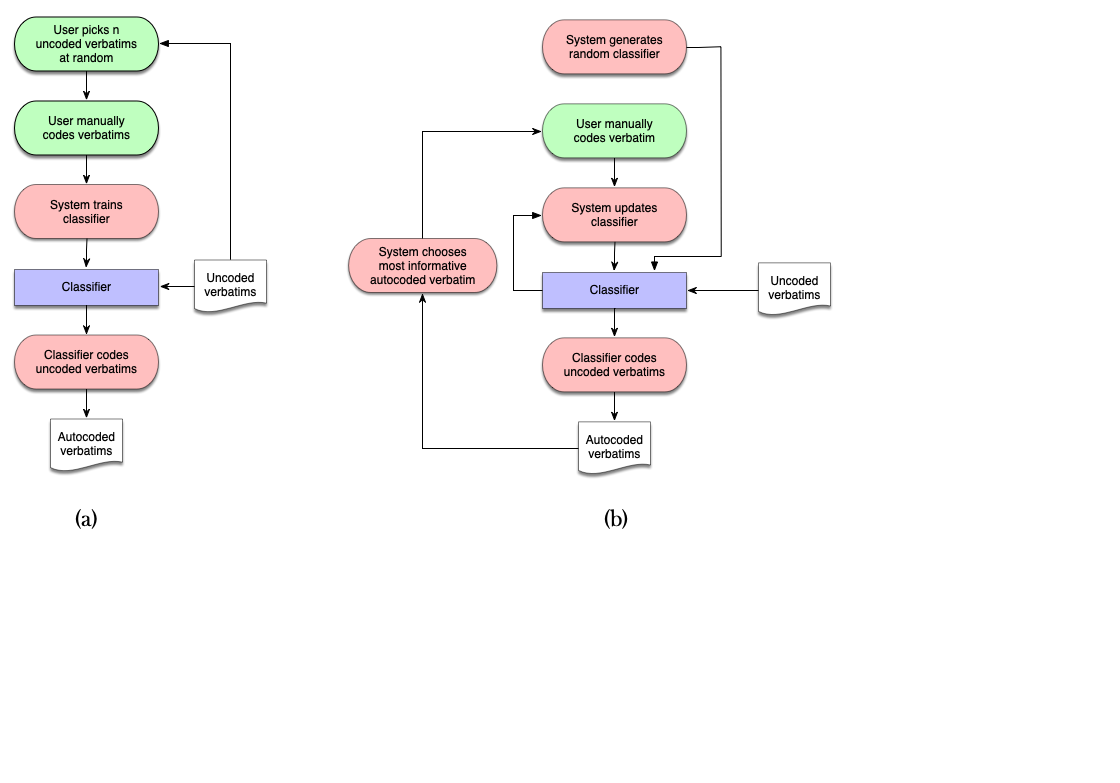}
    \vspace{-30ex}
    \caption{Workflows for automated verbatim coding based on (a)
    standard (i.e., batch passive) learning, and (b) interactive
    (i.e., incremental active) learning. Green
    indicates user actions, while red
    indicates system actions.}
    \label{fig:OldAndNewWisdom}
  \end{center}
\end{figure}

The difference between the ``classic'' verbatim coding systems and the
ones we envisage here is best illustrated in Figure
\ref{fig:OldAndNewWisdom}, where part (a) illustrates the typical
workflow of a classic system and part (b) illustrates instead the
workflow of an interactive-learning-based system.

Let us look at Figure \ref{fig:OldAndNewWisdom}(b) in more detail. The
process starts with
the system generating a random classifier.  This classifier is used to
autocode all the uncoded verbatims, after which the system chooses
(according to a policy that we will discuss), among all the autocoded
verbatims, the one that it expects to be, once validated, the most
informative (when used as a training example) for the training
process. The chosen verbatim is fed to the user to validate, after
which the cycle starts again. \blue{The net effect is that control
passes from user to system and from system to user each time a new
verbatim is validated; this frequent switch, with user and system
working in tightly-bound mode, justifies the name of ``interactive
learning''.}

Therefore, at each iteration, 
(i) the verbatim that the user should validate is chosen by the system
and not by the user, (ii)
the classifier is immediately updated (rather than retrained from
scratch) to reflect the contribution of the new training verbatim, and
(iii) all the uncoded verbatims (with the exception of those which
have been validated already) are autocoded.

The user terminates the loop when the annotation budget is over, or
when
she believes that the trained classifier has become accurate enough.

Note that all the uncoded verbatims that have not been validated

\begin{itemize}

\item are autocoded over and over again, at each iteration. While some
  of them will be assigned the same code in two or more successive
  iterations, some others might not, due to the fact that the
  classifiers employed in the different iterations are different, with
  the more recent one usually being more accurate;

\item are evaluated over and over again, at each iteration, for
  allowing the system to choose the one to be fed to the user for
  validation. Again, a verbatim that has not been deemed useful in a
  previous iteration might be deemed useful in a subsequent one, since
  the classifiers that need to be improved upon are different.
    
\end{itemize}
%


\subsection{The rationale for interactive learning}
\label{sec:rationaleforinteractivelearning}

\noindent What is the rationale for switching from batch passive
learning to incremental active (i.e., ``interactive'') learning? The
main reason behind our proposal is that \emph{not all training items
are created equal}. In other words, assuming we can choose, from a
pool of available verbatims, a subset of $n$ verbatims to be used as
training data, different subsets will lead to classifiers
characterized by different levels of accuracy. AL is a set of
technologies concerned with making an \emph{informed} such choice,
i.e., one that results in classifiers as accurate as
possible. Therefore, the main rationale behind our move is to have
such choice be performed by the system (which we can instruct to make
informed choices) instead of the user (whom we cannot assume to be
able to make informed choices). The higher accuracy that AL can bring
about thus \emph{justifies switching from PL to AL}.

When using AL, the classifier is retrained every time $k$ new training
items are available, and all AL techniques allow setting a desired
value for $k$. Which value of $k$ is the best? Experiments in
different applicative scenarios -- see also Section \ref{sec:results}
-- indicate that \emph{the smaller the value of $k$, the better the
resulting accuracy}. To see why this is the case, assume we have
chosen $k$ to be 10: while choosing 10 informative verbatims from our
pool (and having them validated) may indeed improve accuracy, it is
intuitive that an even higher improvement can be obtained by first
choosing 5 of them, retraining, and \emph{only then} choosing the next
5, since the choice of these latter \blue{\emph{can be informed by the
knowledge of the effects that the previous 5 have brought about.}}
\blue{For instance, when using $k=10$ we might not realize that some
information contained in the first 5 chosen verbatims is duplicated,
or nearly duplicated, by information contained in the last 5. If we
instead choose $k=5$, we retrain after the first 5 verbatims have been
validated, and the 5 verbatims chosen in the next round will likely
not include duplicates of the previous 5.}

This line of reasoning implies that the best possible value of $k$ is
1, i.e., retraining is performed every time a new training verbatim
becomes available.  However, when using traditional batch learning,
the problem with using small values of $k$ is the computational load
involved, since in batch learning the computational cost of training
is a linear (or supra-linear) function of the number of training
examples: just picture the case of having a classifier trained on
1,000 training examples, and (assuming we have chosen $k=5$)
retraining it \emph{from scratch} with 1,005 examples, with 1,010
examples, etc. In incremental learning, instead, a classifier is not
retrained from scratch, but simply \emph{updated} with the information
provided by one single additional training example; in the example
above, updating a classifier trained on 1,000 training examples, thus
bringing to bear the 1,001st example, requires a tiny fraction of the
time that retraining from scratch on 1,001 examples would require,
\blue{because the original 1,000 examples are no more involved}. The
desirability of using $k=1$ in an active learning context thus
\emph{justifies switching from batch learning to incremental
learning}.


\subsection{Choosing the most informative verbatim}
\label{sec:rightverbatim}

\noindent The last important aspect we want to touch upon is the
policy according to which the autocoded verbatim is chosen (for the
user to validate and for the learning algorithm to subsequently use as
a training example).  While many different policies might be
concocted, we here focus on discussing three simple ones:

\begin{itemize}

\item \textsc{Random}: we choose a random verbatim from the pool.

\item \textsc{MinMax}: on odd-numbered (resp., even-numbered)
  iterations we choose the verbatim which the classifier is most
  certain to be (resp., not to be) an instance of the code.
 
\item \textsc{Uncertain}: we choose the verbatim for which the
  classifier is most uncertain whether it should be attributed the
  code or not.

\end{itemize}

\noindent The \textsc{Random} policy is not a policy we might
realistically consider, and we will only use it for comparison
purposes. The other two policies are based on the fact that modern
classifiers, when coding an uncoded verbatim, output not only a binary
decision (indicating whether to attribute the code to the verbatim or
not), but also a confidence score, indicating how certain the
classifier is of its decision; usually, a score of 1 indicates total
certainty that the verbatim should be attributed the code, a score of
0 indicates total certainty that the verbatim should \emph{not} be
attributed the code, while a score of 0.5 indicates total uncertainty
as which between the two is the case.

The rationale of the \textsc{MinMax} policy is that, by alternating
between examples likely to belong to the class and ones likely
\emph{not} to belong to the class, we are likely to generate a set of
positive and negative training examples of the code as balanced as
possible, which should result in an accurate classifier.  The
rationale of the \textsc{Uncertain} policy, instead, is that a
verbatim that the current classifier cannot code with confidence is
likely to be informative (once validated by the user), since it is
likely to help the classifier code correctly examples on which the
classifier is currently uncertain about.  In Section
\ref{sec:experiments} we will present the results of experiments in
which we comparatively run the three policies, and which will help us
in clarifying which policy is the best.


\section{Experiments}
\label{sec:experiments}

\noindent In this section we describe the results of experiments in
which we have tested the accuracy and the efficiency of both
traditional (i.e., batch passive) learning and interactive learning on
a number of datasets of manually coded verbatims.

An experiment consists in training the classifier on a subset of the
verbatims in the dataset (which are thus called the ``training
examples''), using the trained classifier in order to code all the
other verbatims in the dataset (the ``test examples''), and evaluating
how accurately the entire set has been coded as a result of this
process. \blue{Here, ``accuracy'' means adherence to the manually
assigned codes, which are assumed to be correct}; see Section
\ref{sec:results} for details on how this adherence is computed.

\blue{All the discussion in this paper, and the experiments we
describe, focuses on the \emph{binary} case, i.e., on training
classifiers that decide whether a item should be assigned or not a
given code. The reason is that some of the datasets we will use are
binary while the other are multi-label multi-class ones, and we have
seen at the beginning of Section \ref{sec:interactive} that MLMC
classification can be recast as binary classification. We do not run
single-label multi-class experiments, one of the reasons being that
SLMC is a rare occurrence in survey coding (almost all cases that
arise in practice are either of the binary or of the MLMC type);
however, everything we say in this paper readily applies to the SLMC
case too.}



\subsection{Datasets}
\label{sec:datasets}

\noindent A \emph{binary dataset} is a set of texts manually coded to
indicate whether, for a given code $c$, they belong to $c$ or not. A
\emph{MLMC dataset} is instead a set of texts manually coded to
indicate whether, for each code $c$ in a codeframe $\mathcal{C}$ consisting of $m>1$ codes, they belong to $c$ or not. As already discussed in
Section \ref{sec:interactive}, working on a MLMC dataset characterized
by a codeframe with $m$ codes is equivalent to working on $m$ binary
datasets, since multi-label classification is accomplished by
deploying $m$ independent binary classifiers, one for each code in the
codeframe.

In our experiments we use the following datasets, or groups
thereof\footnote{Groups \ref{item:LL-ACE}, \ref{item:LL-BDFGHIL}, and
\ref{item:ANES-L/D}, were originally described and used in
\cite{Esuli:2010kx}, and are also used in \cite{Berardi:2014ys}. Of
group \ref{item:Egg}, the \{EggB1,EggB2\} datasets are the same as
described in \cite{Berardi:2014ys}, while \{EggA1,EggA2\} are random
subsets of the datasets of the same name used in
\cite{Berardi:2014ys}. The reason for taking \emph{subsets} of the
original \{EggA1,EggA2\} is to have all datasets in the same group
contain the same number of verbatims;.}:

\newcounter{GroupsOfDatasets} \setcounter{GroupsOfDatasets}{0}

\begin{enumerate}

  \setcounter{enumi}{\value{GroupsOfDatasets}}
  
\item \label{item:LL-ACE} \textbf{LL-ACE}: this is a set of 3
  multi-label datasets (called LL-A, LL-C, LL-E) of 201 verbatims
  each, returned in response to market research surveys conducted
  around 2009 by Language Logic LLC (now called
  Ascribe\footnote{\url{http://goascribe.com/}}), a US-based company
  specializing in software platforms for market research and active in
  the US since 2000; their codeframes contain 16, 20, 39 codes,
  respectively.\footnote{\label{foot:samenumber}Throughout this
  section, by the number of codes contained in a codeframe we actually
  mean the number of codes in the codeframe \emph{that have at least 1
  instance} in the corresponding dataset; we thus ignore the codes
  that are never instantiated.} \blue{These datasets are from one wave
  of a continuous (``tracking'') survey that the company used to code
  12 times a year, which consisted of ``semi-open'' brand questions
  (i.e., questions -- such as ``What is your favourite soft drink?''
  -- that, although in principle eliciting a textual response, usually
  generate many responses consisting of only the name of a product or
  brand, with this name coming from a small set of such
  names. However, the answers have the typical features of textual
  answers, such as, e.g., different variants of the same brand name
  (e.g., Coca Cola or Coke), padding text (e.g., ``My all-time
  favorite is''), and typos.}

\item \label{item:LL-BDFGHIL} \textbf{LL-BDFGHIL}: this is a set of 7
  multi-label datasets (called LL-B, LL-D, LL-F, etc.) of 501
  verbatims each, returned in response to market research surveys also
  conducted by Language Logic LLC; their codeframes contain 26, 21,
  38, 82, 64, 52, 50 codes, respectively. \blue{These datasets are
  from a large consumer packaged-good study, with both open-ended and
  brand-list questions.}

\item \label{item:Egg} \textbf{Egg}: these are 2 sets of 2 multi-label
  datasets each (\{EggA1, EggA2\} and \{EggB1,EggB2\}); each of the 4
  datasets consists of 926 verbatims returned in response to a
  customer satisfaction survey conducted by Egg
  PLC\footnote{\url{https://www.ybs.co.uk/help/online/egg.html}}, a
  large online bank active in the UK since 1998; the codeframe for
  EggA1 and EggA2 contains 21 codes while the codeframe for EggB1 and
  EggB2 contains 16 codes. \blue{For both sets of datasets, which were
  collected in the context of two different surveys, respondents were
  answering the question ``Have we done anything recently that has
  especially disappointed you?''.}
    
\item \label{item:ANES-L/D} \textbf{ANES-L/D}: this is a set of 2,665
  verbatims returned in response to a political survey conducted
  several years ago by the American National Election Studies
  (ANES)\footnote{\url{http://www.electionstudies.org/}}, a US-based
  project that has been running national surveys of voters since
  1948. \blue{It consists of two mutually disjoint subsets of
  verbatims: the ones in the first subset were returned as an answer
  to the question ``Is there anything in particular about Mr.\ Clinton
  that might make you want to vote for him? If so, what is that?''
  while the ones in the second subset were returned as an answer to
  the question ``Is there anything in particular about Mr.\ Clinton
  that might make you want to vote against him? What is that?''. Our
  coding task consisted in guessing whether the verbatim belongs to
  the first subset (code `Like') or to the second subset (code
  `Dislike').}
    
  \setcounter{GroupsOfDatasets}{\value{enumi}}

\end{enumerate}

\noindent Note that all the datasets in the same group contain the
same number of verbatims. \blue{This is intentional, i.e., we have
grouped the datasets in such a way (see also Footnote
\ref{foot:samenumber}) that this property holds:} since in the rest of
the paper we will sometimes (e.g., in Figure \ref{fig:BestK}) report
the average accuracy of a given system across all datasets in the same
group, this property will prevent us from comparing apples to oranges.

None of the datasets above is publicly available, and we are using
them under tight non-disclosure agreements imposed by the companies /
institutions that own them. Since we know of no publicly available
dataset of answers to open-ended questions, in the interest of
reproducibility we also present results on other datasets that, while
consisting of texts other than answers to open-ended questions, are
nonetheless publicly available. They are\footnote{Each dataset in
group \ref{item:MDS} is a randomly chosen subset of a dataset
available from
\url{https://www.cs.jhu.edu/~mdredze/datasets/sentiment/} (again,
subsets were taken in order to have all datasets in the same group
consist of the same number of items).  Dataset
\ref{item:Reuters-21578(10)} is available from
\url{https://raw.githubusercontent.com/nltk/nltk_data/gh-pages/packages/corpora/reuters.zip}
.}:
\begin{enumerate}
  \setcounter{enumi}{\value{GroupsOfDatasets}}
\item \label{item:MDS} \textbf{MDS}: this is a set of 4 binary
  datasets (called DVDs, Electronics, Kitchen, Books) each consisting
  of 2,000 Amazon product reviews (of DVDs, home electronics items,
  kitchen appliances, and books) coded by sentiment (`Positive' vs.\
  `Negative');
\item \label{item:Reuters-21578(10)} \textbf{Reuters-21578(10)}: this
  is a multi-label dataset of 10,788 Reuters newswire stories, coded
  according to a codeframe consisting of 10 economy-related codes
  \blue{(such as `Earnings', `Acquisitions', etc.). The codeframe of
  the original dataset (known as Reuters-21578) actually contains 115
  codes, but many of them are extremely infrequent (some codes have
  just 1 positive training example); as a result, following many other
  authors in the text classification literature, we here only use the
  10 most frequent codes.}.
  \setcounter{GroupsOfDatasets}{\value{enumi}}
\end{enumerate}
\noindent \blue{Table \ref{tab:datasets} recapitulates the main
characteristics of the 6 groups of datasets.\footnote{\blue{In this
table the ``total number of binary codes'' should be interpreted as
the total number of binary distinctions that need to be captured, or
binary classifiers that need to be trained. This means that, e.g., for
the ANES-L/D dataset this number is just 1, since one just needs a
classifier to tell code `Like' from code `Dislike'; in other words,
the assignment of code `Dislike' should be more properly seen as the
non-assignment of code `Like', i.e., `Dislike' should not be seen as a
proper code. For the very same reason, for group MDS 4 classifiers
need to be generated, i.e., a `Positive' vs.\ `Negative' classifier
for each of the 4 datasets in the group.}}} \blue{Note that, as from
the descriptions above, groups 1, 2, 3, 6 are MLMC datasets and are
about classification by topic, while groups 4 and 5 are binary
datasets and are about classification by sentiment.}

Groups \ref{item:LL-ACE} to \ref{item:Reuters-21578(10)} account for a
total of 497 binary classification experiments, which qualifies as a
fairly substantial experimentation.
\begin{table}[t]
  \caption{\blue{Main characteristics of the 6 groups of datasets we
  use for experimentation. The last 5 columns indicate the number of
  verbatims contained in the dataset, the number of codes the
  codeframe consists of, the average and median length of the verbatim
  (i.e., number of non-unique words contained in it), and the average
  number of codes per verbatim.}}
  {\scriptsize
  \begin{tabular}{|c|c|c|r|r|r|r|r|r|} \hline & \side{\textbf{Group of
    datasets}} & \side{\textbf{Type}} & \side{\textbf{\# verbatims per
    dataset}} & \side{\textbf{Tot \# binary codes}} &
    \side{\textbf{Avg \# words per verbatim}} & \side{\textbf{Median
    \# words per verbatim}\phantom{x}} & \side{\textbf{Avg \# codes
    per verbatim}} & \side{\textbf{Avg \# positive verbatims per
    code}}\\ \hline 1 & LL-ACE & market research & 201 & 75 & 1.80 & 1
    & 1.22 & 9.76 \\ 2 & LL-BDFGHIL & market research & 501 & 333 &
    5.56 & 3 & 1.26 & 13.27 \\ 3 & Egg & customer sat & 926 & 74 &
    26.97 & 22 & 1.74 & 90.60 \\ 4 & ANES-L/D & political survey &
    2,665 & 1 & 26.88 & 21 & 0.52 & 1396.00 \\ 5 & MDS & product
    reviews & 2,000 & 4 & 129.74 & 84 & 0.50 & 1000.00 \\ 6 &
    Reuters-21578(10) & newswires & 10,788 & 10 &
    127.76 & 84 & 0.93 & 997.90 \\
    \hline \multicolumn{3}{c}{\mbox{}}
    & \textbf{Tot} $\rightarrow$ & 497 \\
  \end{tabular}
  }
  \label{tab:datasets}
\end{table}


\subsection{Learning algorithms}
\label{sec:learners}

\noindent As noted at the end of Section \ref{sec:batchvsincremental},
batch learning and incremental learning require different learning
algorithms. For our batch passive learning experiments we have chosen
\emph{Support Vector Machines} (SVMs) \cite{Zhang:2011fl}, a
state-of-the-art learning algorithm that has consistently delivered
top-notch accuracy throughout the text classification
literature\footnote{
For our SVMs we have used a linear kernel since (a) it is the fastest
kernel to train, (b) it has fewer parameters requiring optimization
than other kernels (such as the polynomial or RBF kernels), and (c) it
is the one that usually works best when classifying textual objects,
since text generates very-high-dimensional representations that are
often linearly separable.
}.
 
For our incremental active learning experiments we have instead chosen
an algorithm called \emph{Passive Aggressive} (PA --
\cite{Crammer2006}), also a state-of-the-art incremental learning
algorithm. PA shares many underlying design principles with SVMs, and
(to a first approximation) may be considered as an ``incremental
version of SVMs''; this makes it particularly suitable to our
comparison.

\blue{Note that the specific choice of SVMs and PA to play the roles
of the batch algorithm and the incremental algorithm, respectively, is
inessential to our argument, since what we say in this paper is
largely independent of the specific batch and incremental algorithms
we use.  That is, our argument is that using passive learning with a
batch algorithm $X$ delivers inferior quality with respect to active
learning with an incremental equivalent of $X$; other algorithms we
might have as well chosen to exemplify this could have been, say,
batch \cite{schapire1999improved} and incremental \cite{Oza:2001kc}
versions of AdaBoost.}


%


\subsection{Evaluation measures}
\label{sec:evaluationmeasures}

\noindent As our accuracy measure we use the well-known $F_{1}$
measure (sometimes informally called ``the F-score'', or ``the
F-measure'' -- see \cite{Esuli:2010kx} for a detailed discussion),
defined as
\begin{equation}
  \label{eq:f1}
  F_{1} = \left\{
    \begin{array}{cl}
      \displaystyle\frac{2\cdot TP}{2\cdot TP + FP + FN} & $if$ \ TP + FP + FN>0 \rule[-3ex]{0mm}{7ex} \\
      1 & $if$ \ TP=FP=FN=0 \\
    \end{array}
  \right.
\end{equation}
\noindent where by $TP$, $FP$, $FN$, $TN$, we indicate, as customary,
the number of true positives, false positives, false negatives, true
negatives, obtained by checking the automatically attributed codes
against the true codes (see Table \ref{tab:contingency}); $F_{1}$
values range from 0 (worst) to 1 (best).
\begin{table}[tb]
  \caption{The four-cell contingency table for code $c$ resulting from
  an experiment.}
  \begin{tabular}{|cc|cc|}
    \hline \multicolumn{2}{|c|}{Code} & \multicolumn{2}{c|}{coder says} \\ 
    \multicolumn{2}{|c|}{$c$} &
                                \textbf{YES} & \textbf{NO} \\
    \hline system & \textbf{YES} & $TP$ & $FP$ \\
    says & \textbf{NO} & $FN$ & $TN$ \\ \hline
  \end{tabular}
  \label{tab:contingency}
\end{table}


\subsection{Results}
\label{sec:results}

\begin{figure}[tb]
  \begin{center}
    \includegraphics[width=\textwidth]{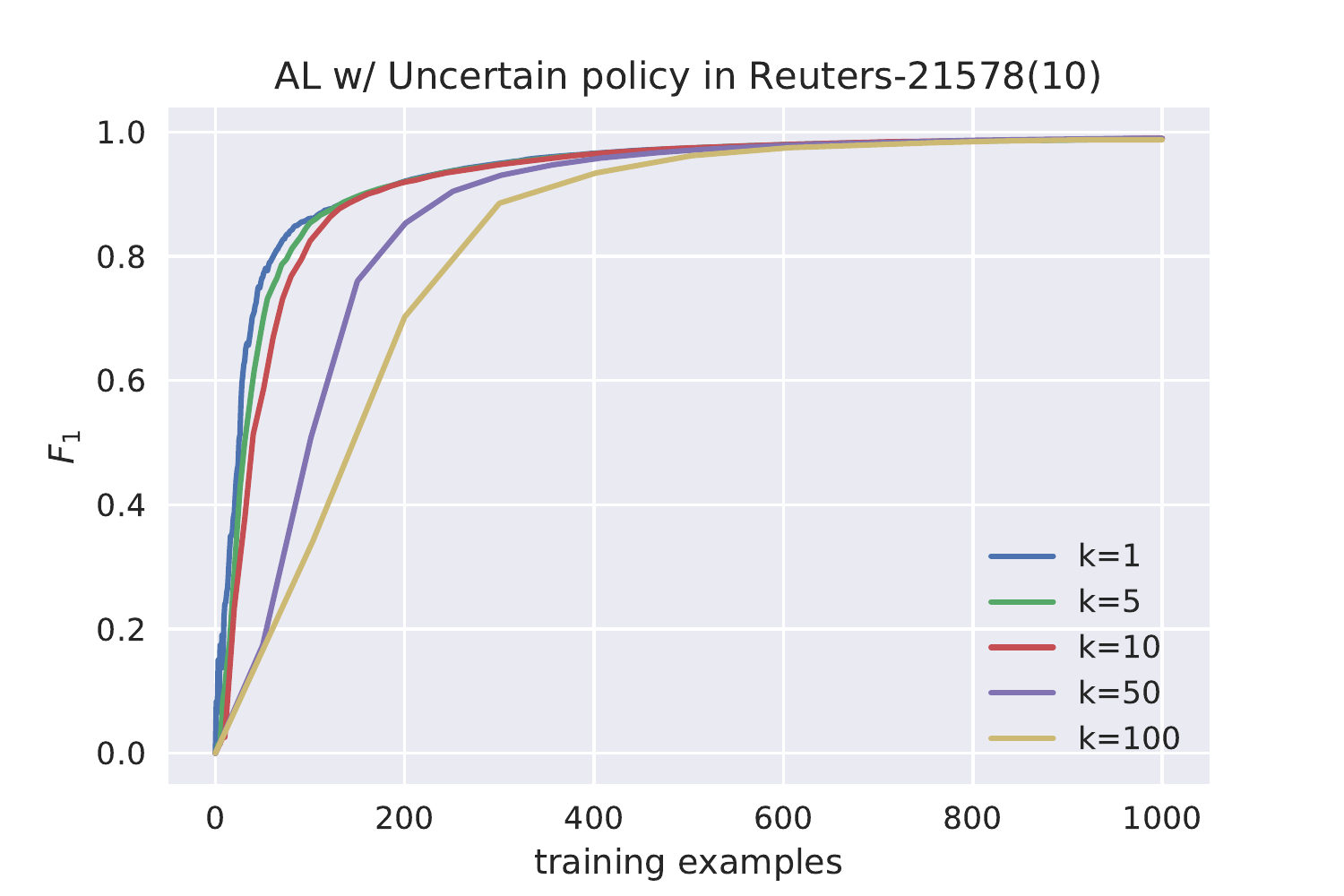}
    \caption{Experiments testing the impact of the value of $k$ on the
    accuracy (measured via $F_{1}$) deriving from active learning
    (here: using the SVM learning algorithm, the \textsc{Uncertain}
    policy, and the Reuters-21578(10) dataset).}
    \label{fig:BestK}
  \end{center}
\end{figure}

\noindent Figure \ref{fig:BestK} represents the results of a number of
experiments that we will discuss more in depth in Section
\ref{sec:bestvalueofk}; for the moment being this figure will only
serve the purpose of illustrating how we display the results of
experiments.
Each among the 5 plotted curves represents the accuracy (expressed in
terms of the $F_{1}$ function of Equation \ref{eq:f1}) of a given
automated coding system on a given group of datasets (here: the
Reuters-21578(10) group) as a function of the number of training
examples used. Higher curves represent better systems, and the
reported accuracy is the average value of $F_{1}$ across all the
datasets (10, in this case) in the group.
If the dataset consists of $N$ verbatims, each point in the curve
represents the result of training on $X$ verbatims, autocoding the
remaining $(N-X)$, and computing the overall accuracy across all the
$N$ verbatims. This is intended to simulate a real scenario in which a
user is interested in obtaining an accurate coding of a set of $N$
verbatims, and for doing it (a) she manually codes $X$ of them for
training a coding system, and (b) autocodes the remaining $(N-X)$ via
the trained system.

Note that curves tend to be increasing from left to right: this
depends on the facts that (a) the higher the number $X$ of training
examples, the more accurate the trained system tends to be, and this
impacts on how accurately the remaining $(N-X)$ verbatims are coded;
(b) the $X$ verbatims chosen as training examples are manually (thus:
correctly) coded, so the higher their number, the higher their impact
on the accuracy of the entire set. This latter aspect also explains
why accuracy is 1 when $X=N$, for all curves: in this case all
verbatims in the dataset have been \emph{manually} coded, no document
requires autocoding, and accuracy is thus maximum.

Note that all our experiments have a random component, since (a) the
batch passive learning experiments involve the random choice of a set
of $X$ training examples, and (b) the incremental active learning
experiments involve the random choice of the initial classifier.  As a
consequence, we have carried out each of our 497 binary classification
experiments by running 10 different trials for each system setup. The
curve that describes a system setup thus results from averaging across
(a) the 10 curves corresponding to the 10 individual trials, and (b)
the binary codes that compose the codeframe.


\subsubsection{What is the best value of $k$?}
\label{sec:bestvalueofk}

\noindent In this experiment we assume we want to train our classifier
via active learning, and we want to establish what is the value of $k$
that leads to the best accuracy. The plot in Figure \ref{fig:BestK}
reports the results of a sample experiment in which we have tested the
effect of different values of $k$ (namely, those in \{1, 5, 10, 50,
100\}); the learning algorithm used is, of course, SVMs, since PA
cannot use values of $k$ higher than 1.
The sample experiment has been carried out using the
\textsc{Uncertain} policy and the Reuters-21578(10) dataset; other
choices of policy and/or dataset have returned similar
results\footnote{The exception is the \textsc{Random} policy, whose
accuracy is by and large independent of the value of $k$. This should
come to no surprise, since with this policy the classifier does not
influence the choice of verbatims to validate.}, and will not be
explicitly reported here for reasons of space.

The experimental results confirm what we had anticipated in Section
\ref{sec:rationaleforinteractivelearning}, i.e., that the smaller the
value of $k$ is, the better active learning works; the optimal value
of $k$ is thus 1. In the experiments that follow we will thus fix the
value of $k$ to 1; this will prompt us to use an incremental (instead
of a batch) learning algorithm, thus giving rise to what we call
``interactive'' learning.


\subsubsection{What is the best active learning policy?}
\label{sec:bestpolicy}

\begin{figure}[tbh]
  \begin{center}
    \includegraphics[width=\textwidth]{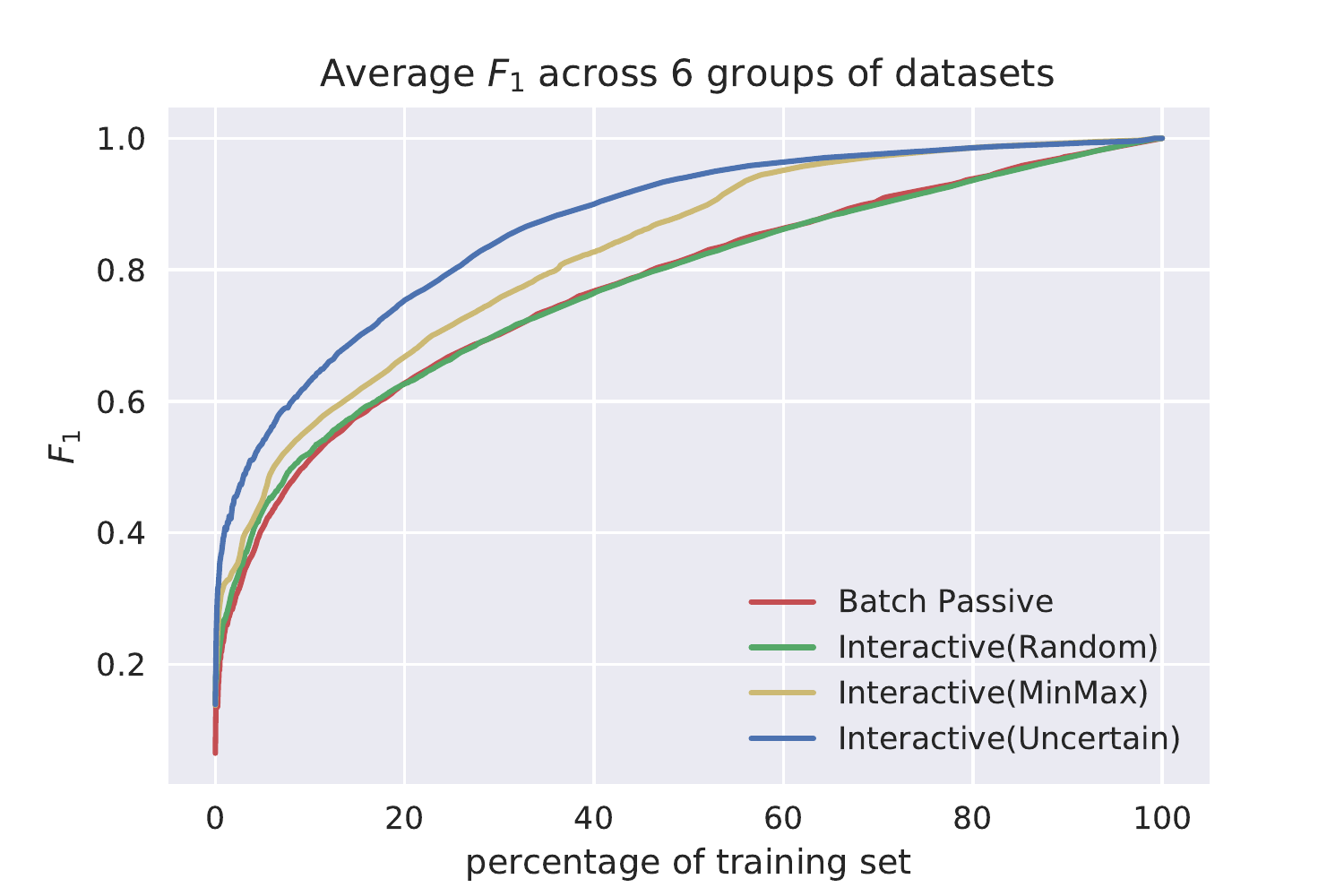}
    \caption{Accuracy, as a function of the percentage of the entire
    dataset that is used as training set, of (a) batch passive
    learning and (b) interactive learning using the 3 different
    policies of Section \ref{sec:rightverbatim}. Each point on a curve
    represents average accuracy across our 497 binary codes $\times$
    10 random trials.}
    \label{fig:BestPolicy}
  \end{center}
\end{figure}

\noindent In order to let us appreciate the relative merits of batch
passive learning and interactive (i.e., incremental active) learning,
Figure \ref{fig:BestPolicy} displays the results of running four
different systems (batch passive learning, plus interactive learning
instantiated with three different policies -- \textsc{Random},
\textsc{MinMax}, \textsc{Uncertain}).
The curves represent $F_{1}$ as a function of the percentage of the
dataset that has been used as training set, and report average $F_{1}$
across all our 6 groups of datasets. Note that, with $k=1$, with batch
passive learning we retrain the classifier \emph{from scratch} every
time a single training example is added to the training set, thus
resulting in a considerable computational load.

There are a few conclusions that we may draw from these results:

\begin{enumerate}

\item In general, \emph{batch passive learning is clearly inferior to
  interactive learning}, since the latter outperforms the former
  across the board (aside from the case in which interactive learning
  is run with the \textsc{Random} policy, which we here only include
  as a reference and not as a serious contender). This means the
  ability, if using interactive learning, to obtain the same accuracy
  as batch passive learning with much less training effort (or: to
  obtain much higher accuracy for the same training effort), and to
  obtain this for any amount of training effort.
\item Batch passive learning has a similar performance as interactive
  learning (to the point that the two curves can be barely
  distinguished) when this latter is run with the \textsc{Random}
  policy. This should come as no surprise, since both systems are
  based on a random (i.e., non informed) choice of training
  examples. When it comes to interactive learning, any policy that has
  a somehow intuitive rationale can be expected to outperform a
  ``non-policy'' such as \textsc{Random}.
\item The \textsc{Uncertain} policy performs substantially better than
  the \textsc{MinMax} policy. This outcome is less obvious, but can be
  explained by the fact that verbatims on which the classifier is
  uncertain are, once validated, very informative, because they help
  the classifier handle the verbatims that it does not classify
  confidently, and would otherwise be most likely to misclassify.  In
  contrast, the verbatims selected by the \textsc{MinMax} policy
  (i.e., those on which the classifier is already confident) are less
  informative, since they merely have the function to reinforce
  beliefs that the classifier already firmly holds. In a sense, by
  being fed the training examples selected by the \textsc{MinMax}
  policy, the classifier ``keeps living in its own bubble''. As a
  recent article titled, ``If you're not outside your comfort zone,
  you won't learn anything''.\footnote{Andy Molinsky, ``If you're not
  outside your comfort zone, you won't learn anything'', \emph{Harvard
  Business Review}, July 29, 2016. \url{http://bit.ly/2ajjIzR}}

\end{enumerate}

\subsubsection{Batch Passive Learning vs.\ Interactive Learning}
\label{sec:batchpassivevsinteractive}

\noindent We have thus determined that, when active learning is used,
(a) $k=1$ is the best setting (see Section \ref{sec:bestvalueofk}) and
(b) \textsc{Uncertain} is the best policy for choosing the verbatim
that the user should validate (see Section \ref{sec:bestpolicy}); in
the interactive learning experiments reported from now on we will
always stick to these two design choices.

\begin{figure}[tbhp]
  \begin{center}
    \includegraphics[width=.49\textwidth]{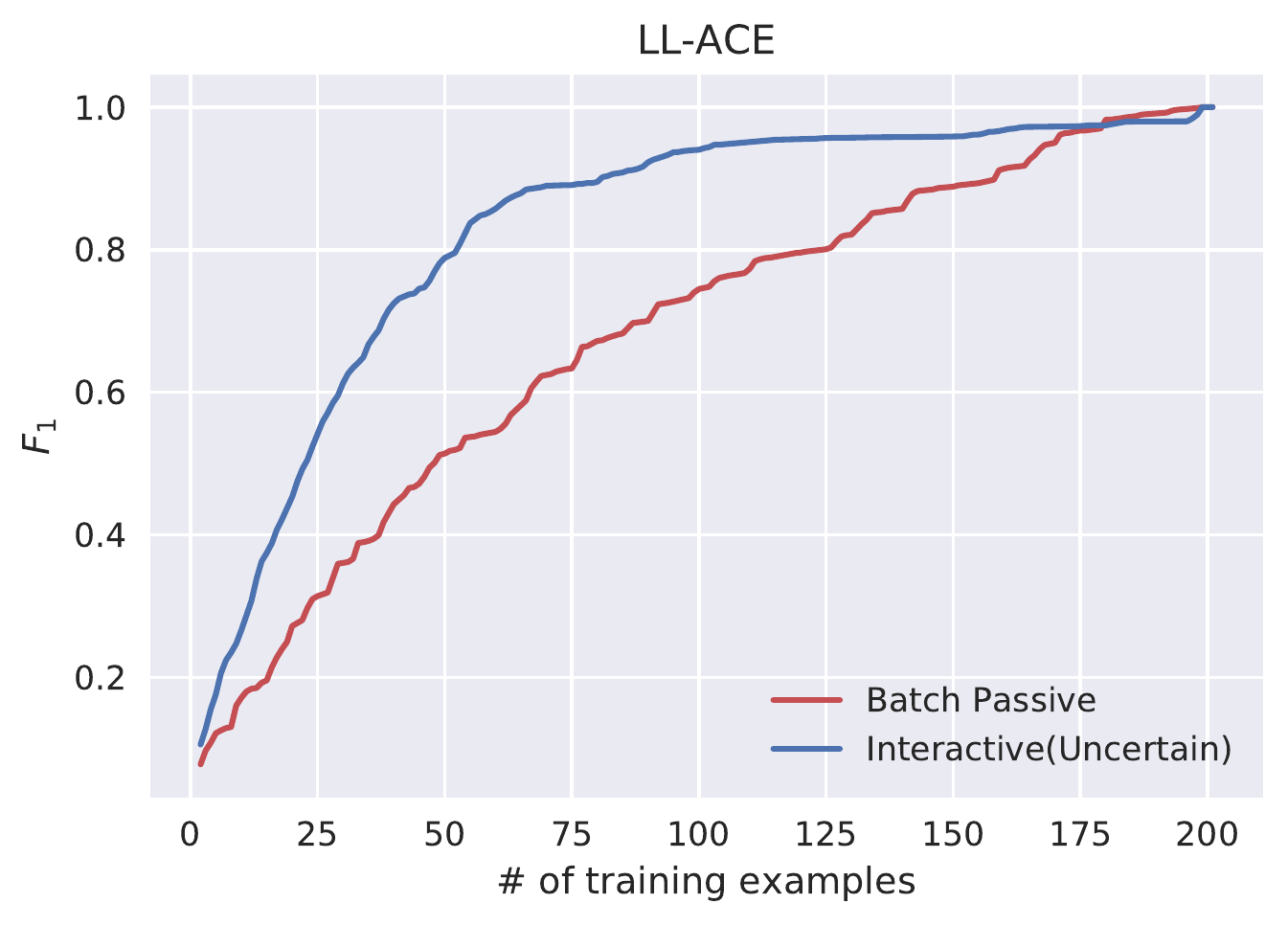}
    \includegraphics[width=.49\textwidth]{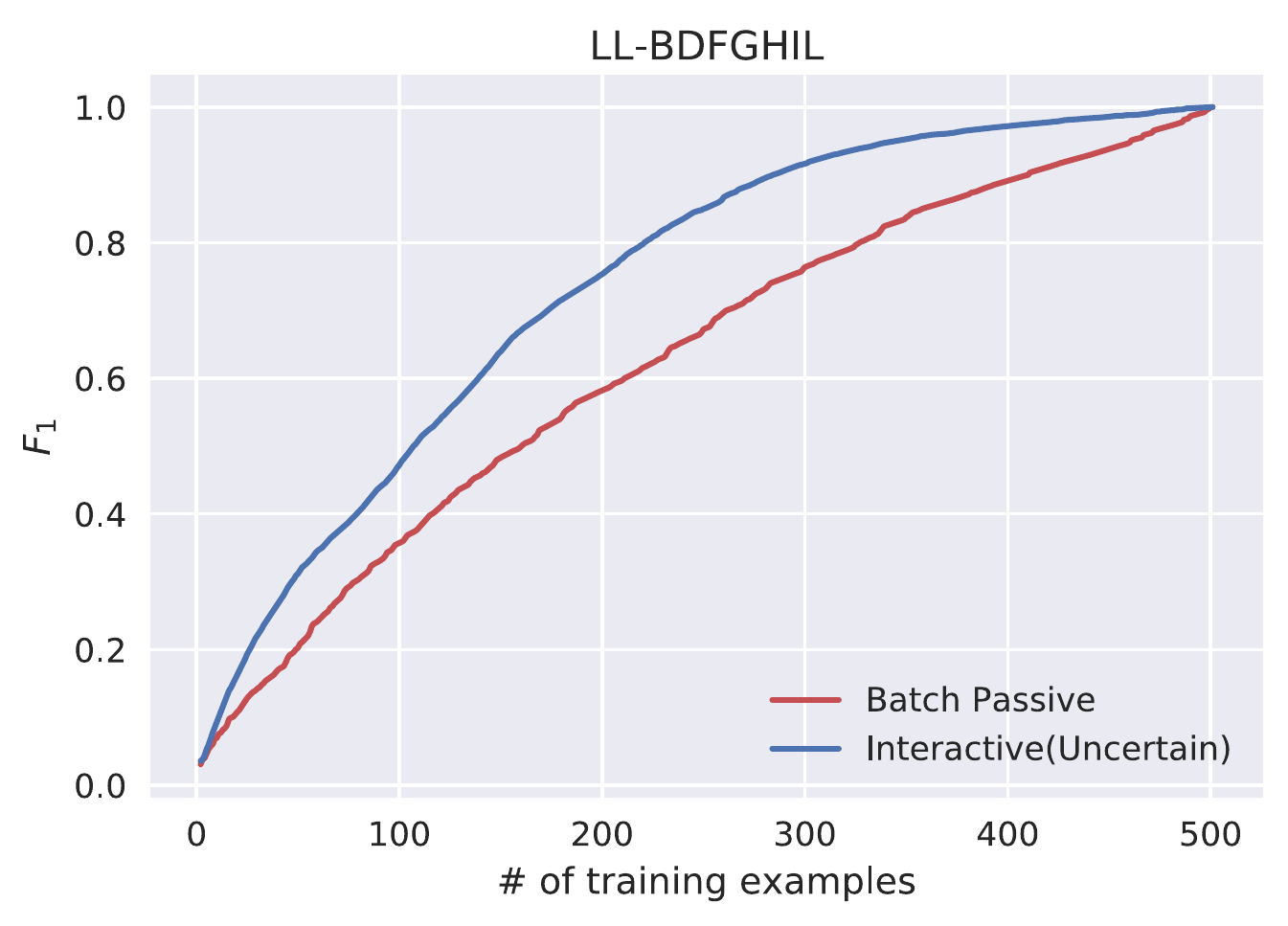}  \\
    \includegraphics[width=.49\textwidth]{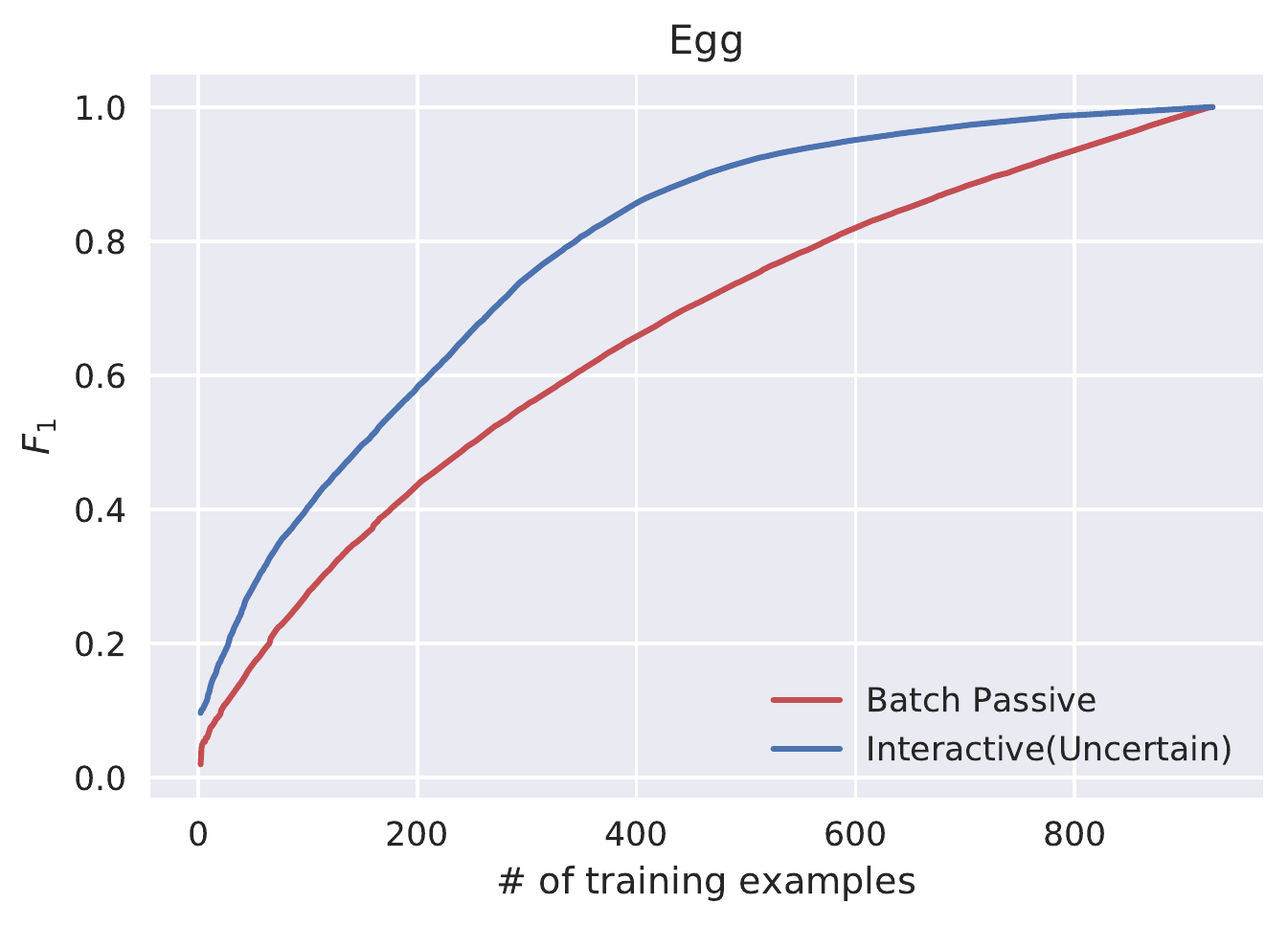}
    \includegraphics[width=.49\textwidth]{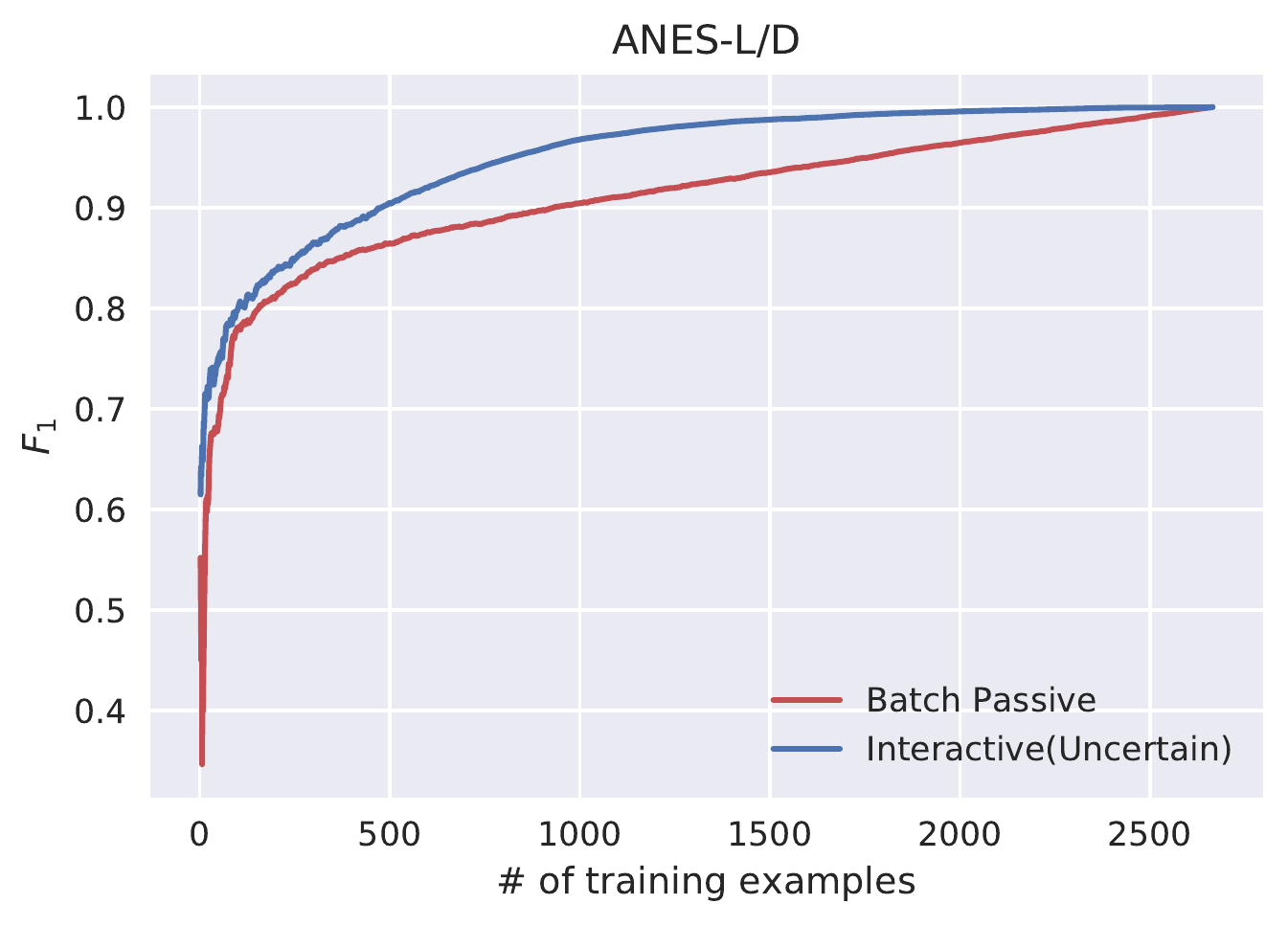}  \\
    \includegraphics[width=.49\textwidth]{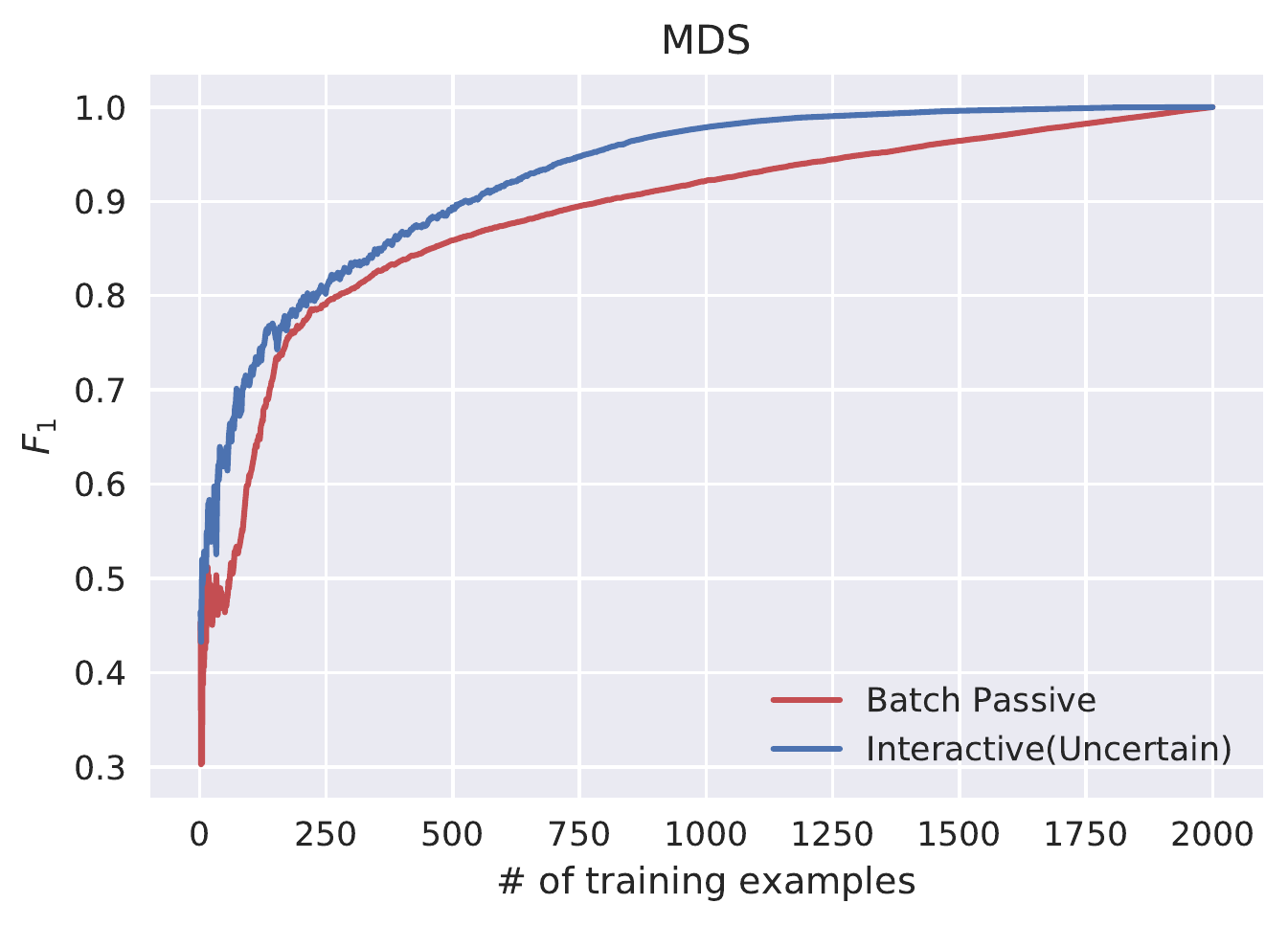}
    \includegraphics[width=.49\textwidth]{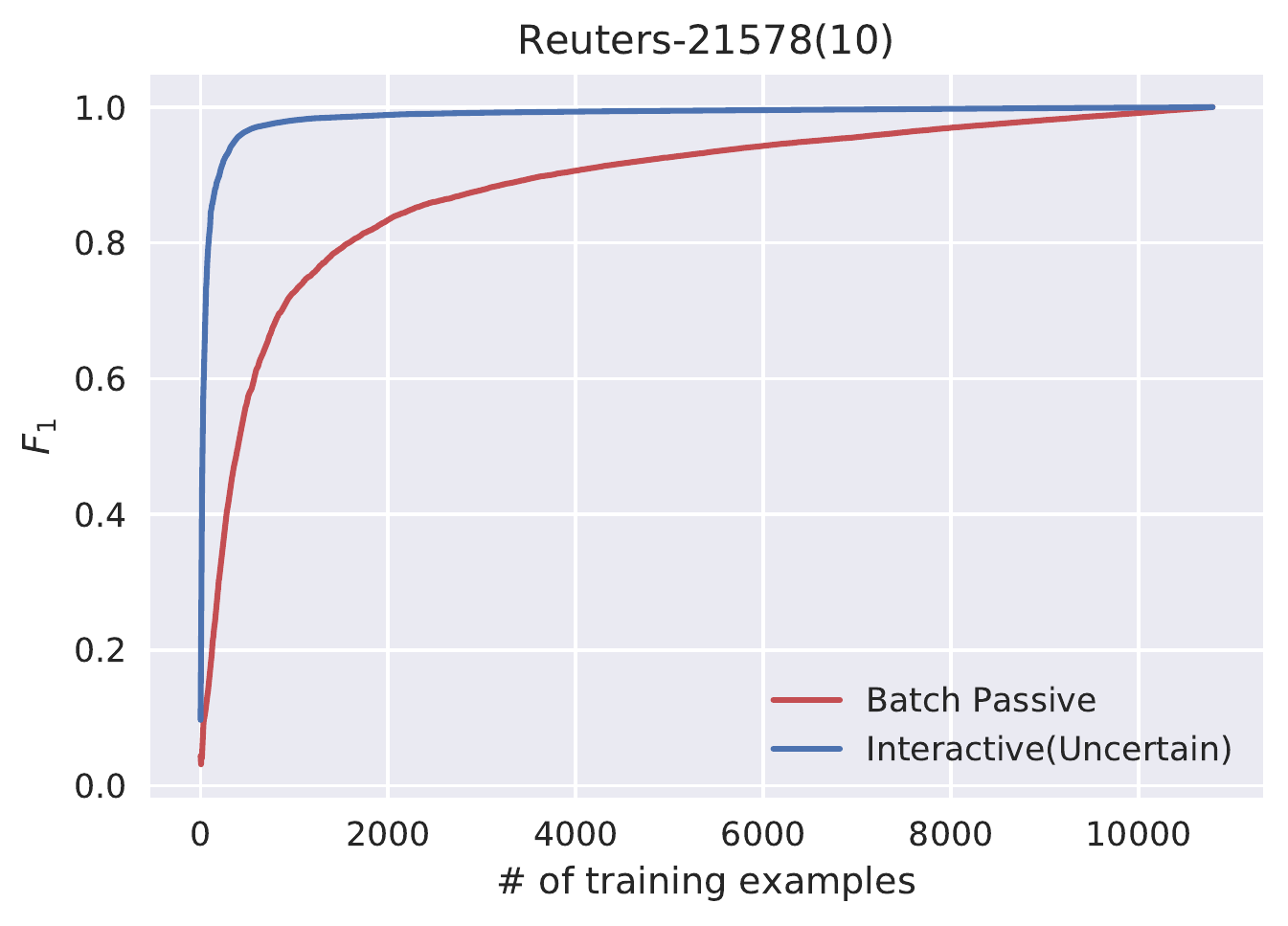}
    \caption{$F_{1}$, as a function of the percentage of the entire
    dataset used as training set, of batch passive learning vs.\
    interactive learning using the \textsc{Uncertain} policy and
    $k=1$. The six figures represent the six groups of datasets
    introduced in Section \protect\ref{sec:datasets}.}
    \label{fig:sixgroups}
  \end{center}
\end{figure}

Figure \ref{fig:sixgroups} experimentally compares batch passive
learning with interactive learning on the six groups of datasets
presented in Section \ref{sec:datasets}.
All of the six subfigures of Figure \ref{fig:sixgroups} unequivocally
confirm the superiority of interactive learning with respect to batch
passive learning, which was the main working hypothesis of this
paper. This means better accuracy with the same amount of training
effort, or less training effort in order to obtain the same level of
accuracy.

Incidentally, this also means that the use of automated verbatim
coding systems based on machine learning may become attractive even
for small-sized studies (i.e., when the number $N$ of verbatims that
require coding is small). For instance, the top left subfigure of
Figure \ref{fig:sixgroups}, which is about the LL-ACE group of
datasets (all consisting of 201 verbatims), shows that manually coding
75 of the 201 examples would result in approximately $F_{1}=0.62$ when
using traditional (i.e., batch passive) learning; in this case, the
effort of generating the training set arguably outweighs the benefits
of having the rest of the dataset autocoded. When using interactive
learning, instead, the same amount of effort leads to approximately
$F_{1}=0.88$, which definitely makes automated coding more attractive.


\subsection{Efficiency}
\label{sec:efficiency}

\noindent An important question that the very notion of interactive
learning raises is that of efficiency, since it is of key importance
that the entire sequence of steps of which one iteration consists
(from user validating a verbatim to user receiving the next verbatim
to validate) can be carried out in real time. Indeed, this problem is
a potential show-stopper, since one iteration requires no less than
(a) the classifier to be updated so as to incorporate the contribution
of the validated verbatim, (b) all the (not-yet-validated) autocoded
verbatims to be autocoded again by the newly updated classifier, and
(c) all these verbatims to be evaluated so that the most promising one
can be singled out for validation by the user in the next iteration.

\begin{figure}[tbh]
  \begin{center}
    \includegraphics[width=\textwidth]{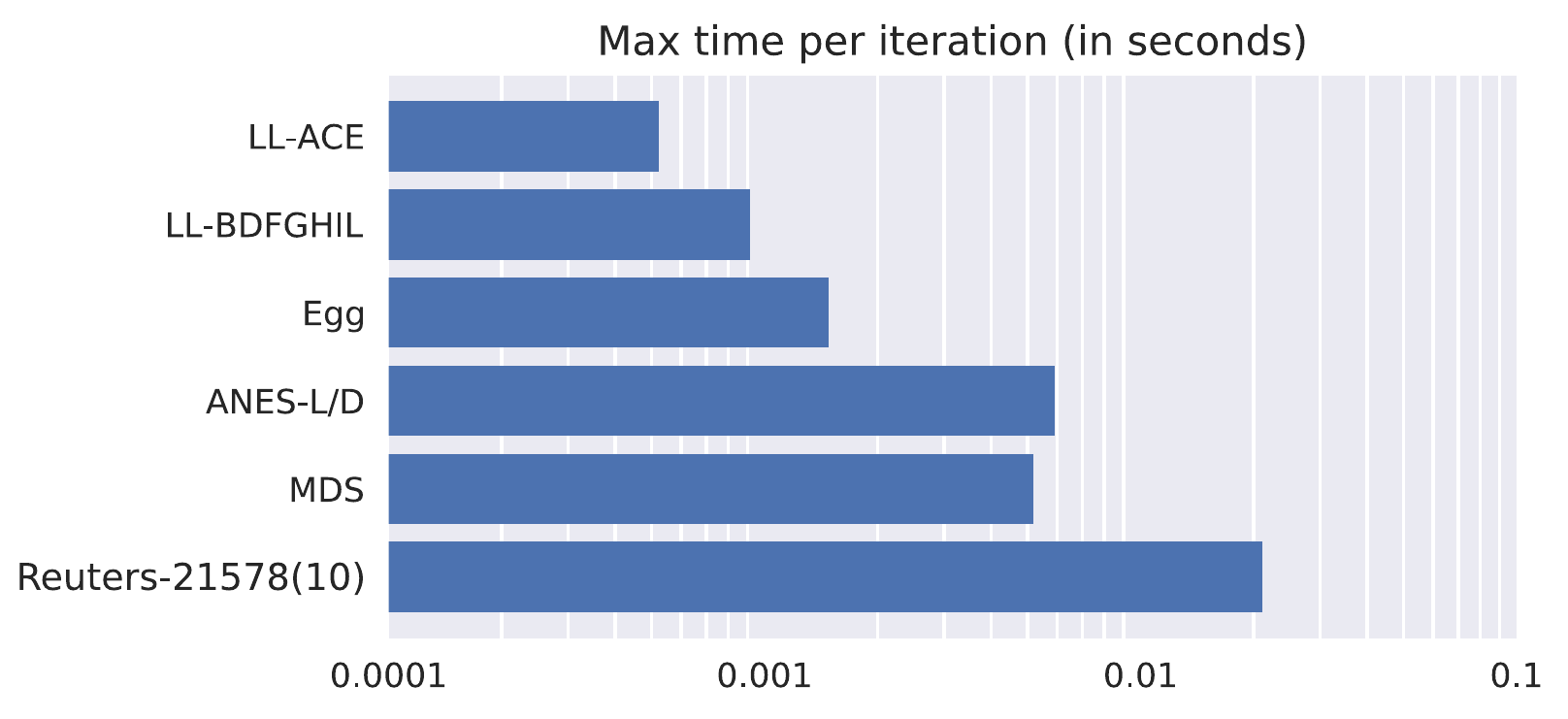}
    \caption{Maximum execution time (in seconds) of a single iteration
    of interactive learning for each of the 6 groups of datasets.}
    \label{fig:times}
  \end{center}
\end{figure}

In order to answer the question above, in Figure \ref{fig:times} we
report, for each among the 6 groups of datasets, the maximum time that
an iteration has requested for any dataset in the group\footnote{The
experiments were run on a commodity machine equipped with an 8-core
processor AMD FX\texttrademark-8350 with 32 GB of RAM under Ubuntu
16.04 (LTS).}.
The histogram shows that execution times vary a lot across groups of
datasets, even by two orders of magnitude; this is intuitive, since
the number of verbatims that need to go through steps (b) and (c) of
the above description varies a lot, from 201 (LL-ACE) to 10,788
(Reuters-21578(10)).

Still, the key observation is that these times are all very low;
thanks to a highly optimized implementation, the \emph{highest}
execution time (required at the very beginning of the process -- when
practically all the verbatims in the dataset need to go thorough steps
(b) and (c) -- in processing Reuters-21578(10), a dataset of 10,788
verbatims) is about 0.02 seconds, which is fast enough for allowing a
smooth interaction between user and machine. \blue{Since computation
times are essentially linear in the number of verbatims that need
coding, this would mean that with a dataset 50 times as large (i.e.,
10,788$\times$50=539,400 verbatims) we would still be able to run each
iteration in under one second.}


\section{Classifier Reuse via Interactive Learning}
\label{sec:reuse}

\noindent The use of interactive learning is also beneficial for the
purpose of reusing classifiers previously trained for different
(albeit related) tasks.
To see how, suppose we need to generate a classifier that codes
verbatims as `Positive' or `Negative' (i.e., a sentiment classifier),
where the verbatims have been returned following a specific question
(hereafter called the \emph{target} question -- say, how the
respondent liked a given restaurant). Suppose we already have (from a
previous study) a classifier that also codes verbatims as `Positive'
or `Negative', but trained on verbatims returned following a different
question (the \emph{source} question -- say, what the respondent
thinks about a given camera).

Given that the codes are the same, should we reuse the ``source''
classifier for coding our ``target'' verbatims? Just relying on the
source classifier seems risky, since the two domains (restaurants and
cameras) are presumably characterized by fairly different ways to
express sentiment. However, since \emph{some} ways of expressing
sentiment can indeed be used for both domains (e.g., adjective
``disastrous'' conveys the same sentiment in both domains), it would
seem attractive to reuse the source classifier and tailor it to the
target question with target-specific training examples, instead of
training a classifier from scratch by just using the target-specific
training examples. But this is something we cannot do if our learning
technology is of the ``batch learning'' type, since for training a
classifier that uses both source and target information we would still
need to access the source training examples, which may not be
available anymore.

If our technology is of the ``interactive learning'' type, though, we
can take the source classifier and incrementally update it by
leveraging the target training examples. In doing so, we may expect
the source classifier to provide an initial, suboptimal solution we
can start from, and we may expect this solution to improve as the
target-specific training examples are employed to refine it.

\begin{figure}[tbh]
  \begin{center}
    \includegraphics[width=.49\textwidth]{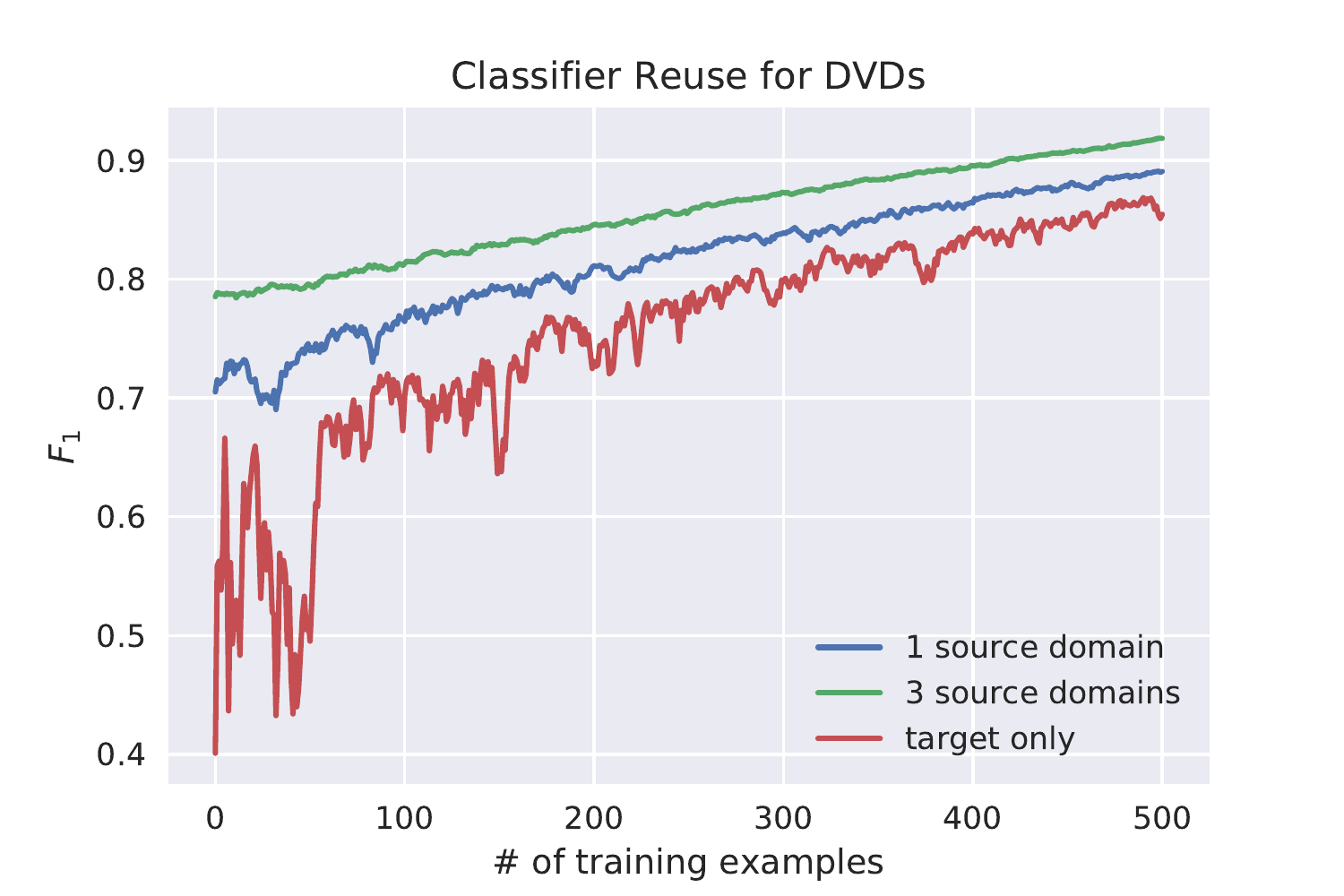}
    \includegraphics[width=.49\textwidth]{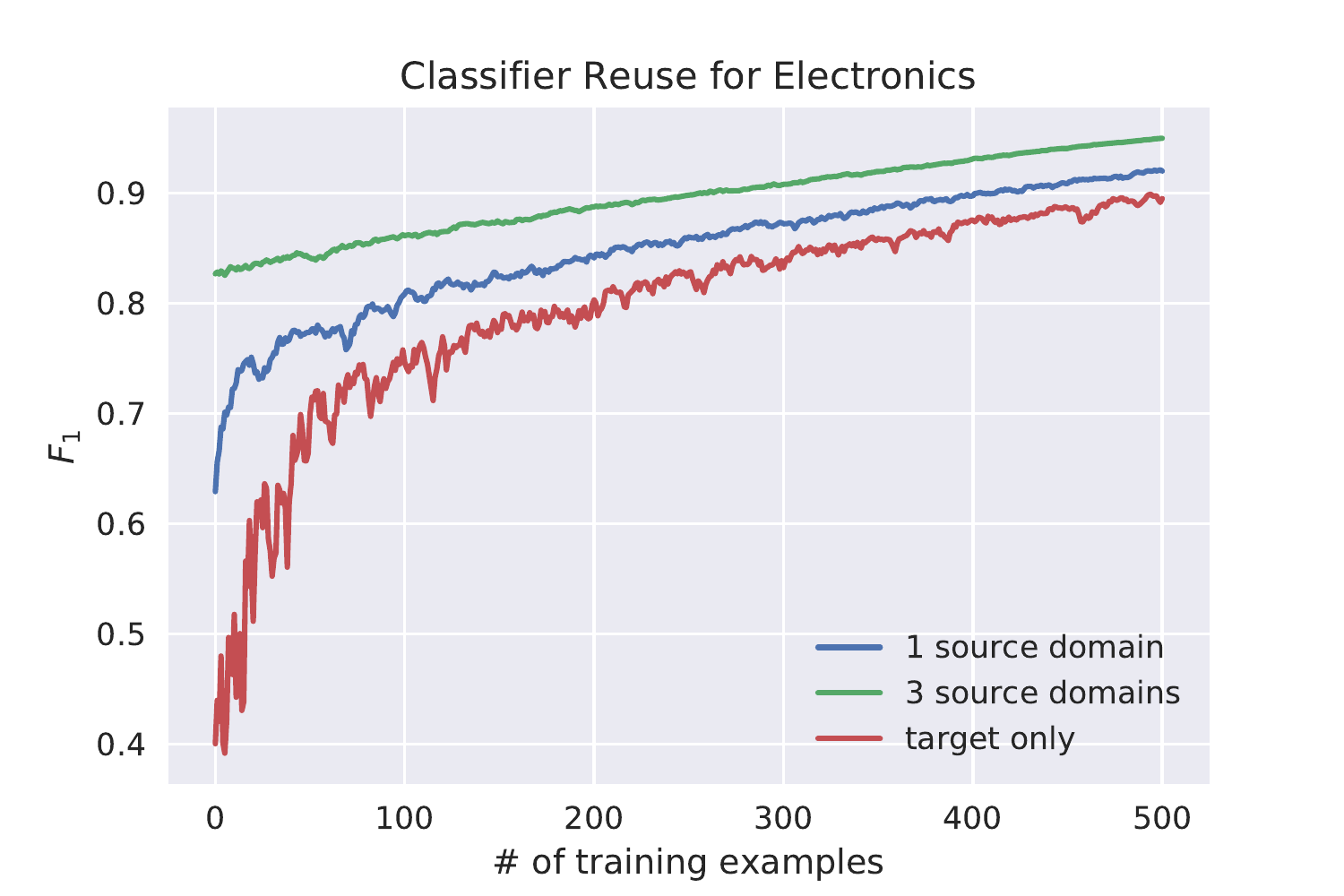} \\
    \includegraphics[width=.49\textwidth]{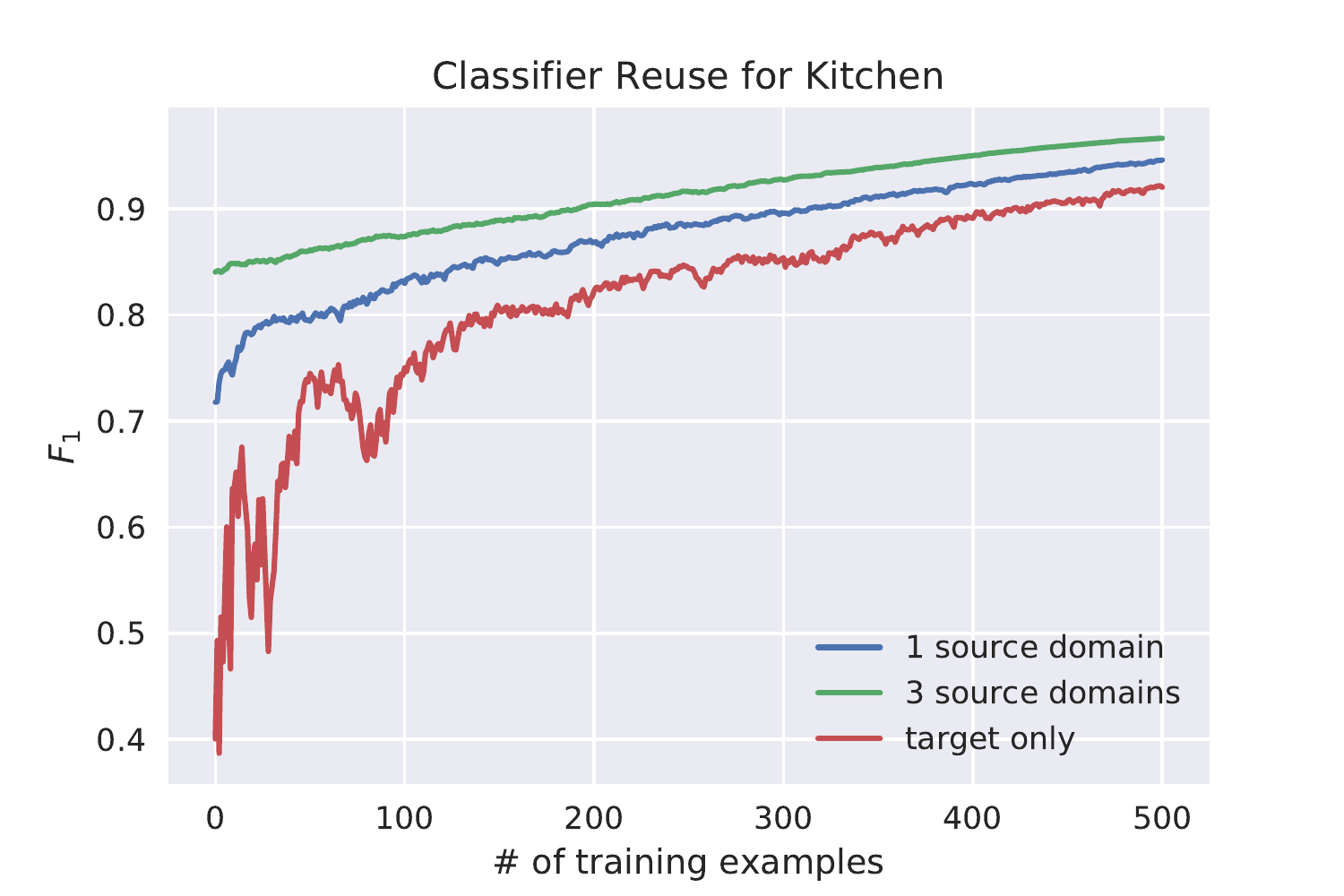}
    \includegraphics[width=.49\textwidth]{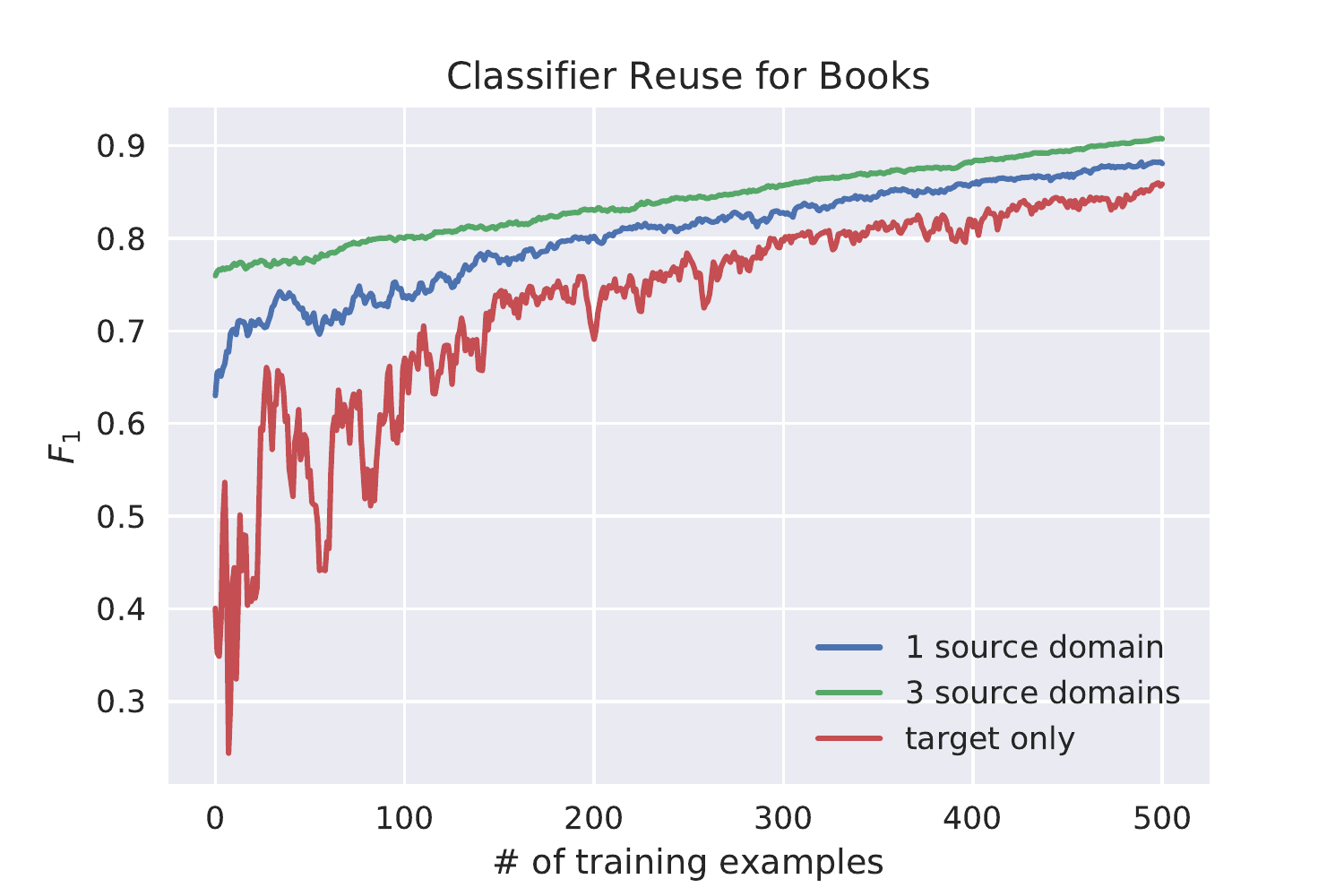}
    \caption{Experiments testing the impact of reusing classifiers
    trained on one or more (``source'') domains for performing
    sentiment classification on another (``target'') domain. The plot
    shows accuracies for the first 500 (out of a total of 2,000)
    target training examples only.}
    \label{fig:transferdvds}
  \end{center}
\end{figure}

In order to check if this approach makes sense, we have run
experiments on the four sentiment classification binary tasks (DVDs,
Electronics, Kitchen, Books) of the MDS group of datasets mentioned in
Section \ref{sec:datasets}; the results are reported in Figure
\ref{fig:transferdvds}.
For instance, in the `DVDs' experiment (top left subfigure), by using
interactive learning we have

\begin{itemize}

\item trained a classifier using only `DVDs' examples, resulting in
  the curve labelled ``target only'';

\item (a) trained three classifiers (one on `Electronics', one on
  `Kitchen', one on `Books' -- each of them using all the available
  verbatims from the respective datasets), (b) updated each of them
  incrementally by bringing to bear, one by one, the `DVDs' examples,
  (c) evaluated the resulting accuracies on the remaining `DVDs'
  examples, and (d) obtained the curve labelled ``1 source domain'' as
  the average of the three resulting curves;

\item (a) trained a classifier on the union of the `Electronics',
  `Kitchen', and `Books' verbatims, (b) updated it incrementally by
  bringing to bear, one by one, the `DVDs' examples, (c) evaluated the
  resulting accuracy on the remaining `DVDs' examples, resulting in
  the curve labelled ``3 source domains''.

\end{itemize}

\begin{figure}[tbh]
  \begin{center}
    \includegraphics[width=1.00\textwidth]{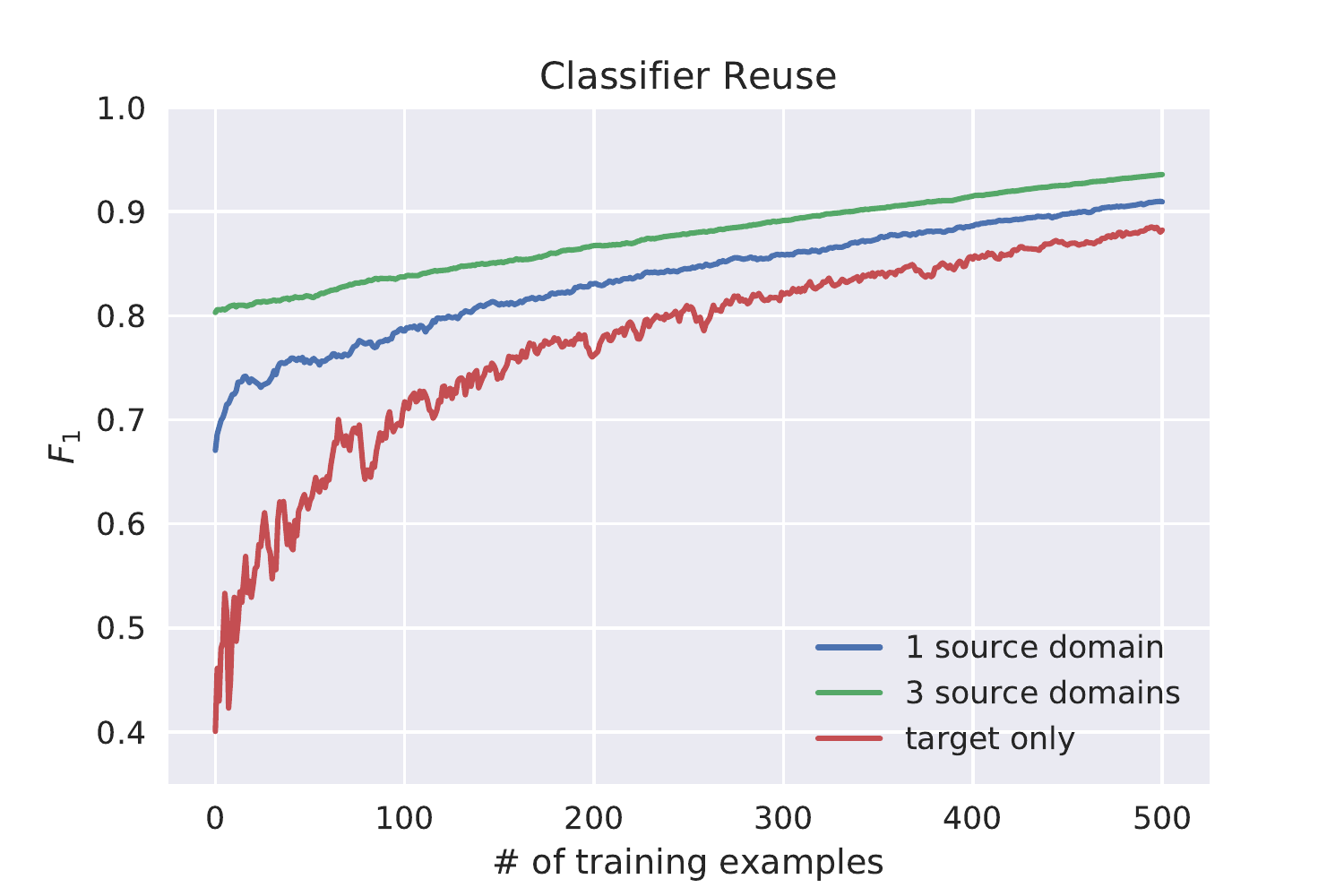} \\
    \caption{Experiments testing the impact of reusing classifiers;
    each curve is the average of the four corresponding curves in
    Figure \ref{fig:transferdvds}.}
    \label{fig:transferavg}
  \end{center}
\end{figure}

\noindent It is immediate to notice from the `DVDs' subfigure of
Figure \ref{fig:transferdvds} that (a) reusing a classifier previously
trained on a different source domain (curve ``1 source domain'') may
be highly beneficial, notwithstanding the semantic difference among
the source and target domains, and (b) reusing a classifier previously
trained on \emph{several} different source domains at the same time
(curve ``3 source domains'') may be even more beneficial. For
instance, if we had to start from scratch by using only
target-specific training examples (curve ``target only''), only by
deploying the first 300 of them we would reach the accuracy that just
using a classifier previously trained on three different source
domains grants us. In other words, the source classifiers give us a
very good base to begin with (with $F_{1}$ accuracy on the target data
between 0.70 and 0.78), and updating them interactively by means of
the target training examples allows accuracy to smoothly and
systematically increase. The accuracy of the system that does not
reuse any previous classifier rises more briskly, but (while getting
close to it) is never able to reach the accuracy of the systems that
do reuse previous classifiers. The `Electronics', `Kitchen', and
`Books' subfigures of Figure \ref{fig:transferdvds} essentially
confirm the intuitions obtained from the results of the `DVDs'
experiment. Figure \ref{fig:transferavg} reports the results of
averaging across the four cases reported in Figure
\ref{fig:transferdvds}.

Note that in order to reuse a classifier in the way exemplified here,
what we only need is that the binary distinctions addressed in the
source and target tasks are the same. While we have here exemplified
this process in the case of classification by sentiment (`Positive'
vs.\ `Negative'), we can do the same for classification by topic. For
instance, the source task could consist in coding whether the reason
for unhappiness of the customers of an online bank is or not the
quality of the website, while the target task could consist of the
same for the customers of a telecom company. We plan to address the
potential of classifier reuse more thoroughly in a future paper.

\section{Related work}
\label{sec:related}

\noindent \textbf{Automated survey coding.} The use of computer
systems that automatically code verbatims is widespread in market
research practice; still, the literature on this topic is extremely
scarce, which unfortunately indicates that little published
experimental evidence exists as to which among the different
approaches on the market is the best. Earlier attempts at automating
the survey coding process were based on classifiers that were not
machine-learned, but human-engineered; examples of this approach are
the systems described in
\cite{Baek:2011vh,Macchia:2002az,Patil:2013zp,Viechnicki98}. Manually
engineering verbatim classifiers is disadvantageous, since manually
writing the classification rules (or the dictionary entries) which are
the essential building blocks of these classifiers is onerous, and is
not necessarily conducive to high classification accuracy. As a
result, the current tendency is to rely on the supervised machine
learning approach.

\medskip\noindent \textbf{Machine learning for automated survey
coding.} The idea to apply supervised machine learning to automating
survey coding was first presented in \cite{ASCIC03,JASIST03}; these
two papers discuss a single-label multi-class survey coding system
based on support vector machines, and present experimental results
obtained on data from the General Social Survey carried out by the US
National Opinion Research Center (NORC). The same research group later
introduced the first commercially available survey coding system based
on machine learning, and discussed its application to coding verbatim
answers obtained as a result of customer satisfaction surveys
\cite{MRS07}, market research surveys, or political surveys
\cite{Esuli:2010kx}. Systems along the lines of the one of
\cite{JASIST03} are described in
\cite{Clarke:2011os,Gamon:2004:SCC,Mantecon:2018pi,Spasic:2018qf}.

In \cite{Berardi:2014ys}, the authors introduce a ``semi-automated
coding'' system, i.e., a system in which coding by a machine-learned
classifier is followed by a phase in which a user validates the
verbatims that, when validated, bring about the highest expected
improvement in the overall accuracy of the entire set; a
utility-theoretic framework tailored on the adopted accuracy measure
is employed to determine these expected improvements. Semi-automated
coding is also discussed in \cite{Schonlau:2016if}; however, these
authors rely on the user to decide the threshold that separates the
verbatims that should be validated from the ones which should not, and
this may exceed what should reasonably be expected from users. A
semi-automated approach analogous to the latter is applied to
occupation coding in \cite{Schierholz:2014tx}.

All of the above works employ a traditional ``batch learning'' method,
i.e., one in which the classifier is generated non-incrementally (the
entire set of training examples are used in one shot) and via
``passive learning'' (i.e., the verbatims are chosen by the user, and
not by the system).

\medskip\noindent \textbf{Active learning and incremental learning.}
While the joint application of active learning and incremental
learning in a survey coding scenario is new, neither active learning
nor incremental learning \textit{per se} are new in the field of
machine learning. In particular, active learning goes back at least to
\cite{Angluin:1988ek}, and both the \textsc{Uncertain} and
\textsc{MinMax} policy can be traced back to \cite{Lewis94a};
incremental learning, instead, even goes back to \cite{Ros58}.

The use of active learning in automated survey coding was first
proposed in \cite{Esuli:2010kx}. Differently from the present paper,
and due to the use of a batch learning algorithm, in that work
classifier retraining is performed anew every time an additional batch
of $k$ autocoded verbatims (instead of just a single autocoded
verbatim) have been validated by the user. Additionally, in
\cite{Esuli:2010kx} active learning is used to improve a classifier
originally trained on a batch of (randomly selected) verbatims, and
not to generate the classifier right from the start. Another work in
which active learning is used to improve a classifier (previously
generated via batch learning) for responses to open-ended questions is
\cite{Patil:2015nk}.

We are not aware, instead, of any previous attempt to use incremental
learning for survey coding applications.


\section{Discussion and conclusions}
\label{sec:discussion}

\noindent When building an automatic verbatim coding system,
interactive learning provides several advantages with respect to the
``classic'' learning metaphor based on passive learning + batch
learning.


The use of active learning (instead of passive learning), i.e., the
idea that it is the system's (instead of the user's) responsibility to
decide which verbatims should be manually coded in order for them to
be used as training data, delivers (for the same amount of training
effort) substantially better accuracy, since the system can ask the
user to manually code exactly those verbatims that, when used as
additional training data, are estimated to provide the highest benefit
to classifier retraining.

We have seen that the increase in accuracy (with respect to
non-interactive learning systems) is especially high when the number
of training items is low. Aside from showing that this allows
automatic classifiers to perform respectably just after a few training
examples have been provided, this also means that interactive learning
is especially attractive for studies in which the annotation budget is
low. This is the case, for instance, of small studies, since in
studies characterized by a low number of verbatims that require
coding, the number of training verbatims that are needed to bring
about sufficient coding accuracy may be close (if ``classic'' systems
are used) to the effort needed to manually code the entire set. In
other words, interactive learning has the potential to make
machine-learned classifiers amenable to dealing also with small
studies, which is not the case in the realm of batch passive learning.

The accuracy of classifiers trained via active learning increases,
quite obviously, with the frequency of retraining operations. Most
active learning literature assumes that retraining is performed every
time $k$ new training examples are available, with $k$ usually in the
dozens or in the hundreds; retraining every time a single new training
example is available is the ideal situation for active learning (we
might view this as a form of ``extreme active learning''), but incurs
severe computational costs when traditional ``batch learning''
technology (that retrains the classifier anew from the entire training
set) is used. The adoption of incremental learning technology makes it
straightforward to retrain every time a new training item is
available, since incremental learning is exactly about updating a
previously trained classifier by bringing to bear a single new
training example. \blue{Our experiments, which we have carried out
across a set of binary datasets and multi-label multi-class datasets,
have shown consistent and substantive accuracy improvements with
respect to ``standard'' batch passive learning. While we have not run
tests on single-label multi-class classification, everything we have
said about interactive learning straightforwardly extends to it.}


Interactive learning can be efficient: we have shown that the entire
cycle triggered by the validation of the chosen autocoded example
(i.e., ``update classifier'' $\rightarrow$ ``autocode all uncoded
verbatims'' $\rightarrow$ ``choose autocoded verbatim for the user to
validate'') can be performed (on datasets of several thousands
verbatims) in a fraction of a second, thus allowing the user to carry
on her validation activity smoothly, and without even realizing that
uncoded verbatims are all re-coded every time she validates a
verbatim. The result is a ``train-while-u-code'' system, i.e., a
learning process that integrates much better in the typical workflow
of the survey specialist, which may perform her manual coding activity
without even realizing that a machine learning process is
ongoing. Rather than a ``human in the loop'' computerized system,
interactive learning implements a ``machine in the loop'' workflow for
the survey specialist, who thus remains at center stage.


\section*{Acknowledgements} \noindent Most of this material was
originally presented in a talk given by the third author at a
conference organized by the Association of Survey Computing, London,
UK, November 2017. Thanks to Ivano Luberti for several interesting
discussions on the topic of this paper.


\bibliography{Fabrizio} \bibliographystyle{klunamed}

\end{document}